\documentclass[final]{svjour3} 
\usepackage{graphicx}
\usepackage{epstopdf}
\usepackage{lscape} 
\usepackage{txfonts}
\usepackage{natbib} 
\bibpunct{(}{)}{;}{a}{}{,}

\newcommand{\apjl}{ApJL}
\newcommand{\apjs}{ApJS}

\newcommand{\aap}{A\&A}

\usepackage{lscape}

\usepackage[usenames]{color} 
\definecolor{MyRed}{rgb}{0.9,0.0,0.0} 
\definecolor{MyLightRed}{rgb}{1.0,0.0,0.0} 
\definecolor{MyPink}{rgb}{1.0,0.08,0.45} 
\definecolor{MyDarkBlue}{rgb}{0,0.08,0.45} 
\definecolor{MyDarkGreen}{rgb}{0,0.5,0.0} 
\newcommand{\edt}[1]{{\textcolor{blue}{#1}}}
\renewcommand{\edt}[1]{#1}

\begin{document}
\title{Coupling from the photosphere to the chromosphere and the corona}

\author{S.~Wedemeyer-B\"ohm, A.~Lagg, {\AA}.~Nordlund}
\institute{
   Sven Wedemeyer-B\"ohm \at
   Institute of Theoretical Astrophysics, University of Oslo, Norway,
            \email{svenwe@astro.uio.no}
\and
   Andreas Lagg \at 
   MPI f{\"u}r Sonnensystemforschung, 
   Katlenburg-Lindau, Germany, 
            \email{lagg@mps.mpg.de}
\and 
   {\AA}ke~Nordlund \at
   Niels Bohr Institute, University of Copenhagen, Denmark,
            \email{aake@nbi.dk}
}
\thanks{
Marie Curie Intra-European Fellowship of the European Commission 
  (6th Framework Programme, FP6-2005-Mobility-5, Proposal No. 042049).   
}
\maketitle

\begin{abstract}
The atmosphere of the Sun is characterized by a complex interplay of 
competing physical processes: convection, radiation, 
conduction, and magnetic fields.
The most obvious imprint of the solar convection and its overshooting
in the low atmosphere is the granulation pattern.
Beside this dominating scale there is a more or less smooth
distribution of spatial scales, both towards smaller and larger scales, 
making the Sun essentially a multi-scale
object.  Convection and overshooting give the photosphere its face but also
act as drivers for the layers above, namely the chromosphere and corona.  The
magnetic field configuration effectively couples the atmospheric layers on
a multitude of spatial scales, for instance in the form of loops that are anchored in
the convection zone and continue through the atmosphere up into the chromosphere
and corona.
The magnetic field is also an important structuring agent for the small,
granulation-size scales, although (hydrodynamic) shock waves also play an 
important role --- especially in the internetwork atmosphere where mostly weak fields
prevail.  Based on recent results from observations and numerical simulations,
we attempt to present a comprehensive picture of the atmosphere of the quiet Sun
as a highly intermittent and dynamic system.
  
\end{abstract}

\section{Introduction}

Observations of the solar atmosphere reveal a wealth of different phenomena,
which occur over an extended range of different temporal and spatial scales.
This is not surprising, considering the fact that already basic parameters
such as gas density and temperature span many orders of magnitude, from the
convection zone below the photosphere to the corona.  At a first look, it may
thus appear rather hopeless to construct an overall picture that can account for
all the phenomena.  At a closer look, however, many connections between
apparently independent phenomena can be found, ultimately implying a multitude
of couplings through the atmosphere.  In addition, there seems to be a 
hierarchical arrangement of approximately selfsimilar convective motions, 
with the granulation pattern embedded in increasingly larger meso- and 
supergranulation patterns.

The key to a comprehensive picture of the solar atmosphere thus lies in relaxing 
too strict and oversimplified concepts, even when they are didactically nicer than
the reality.  The solar atmosphere should not be seen as a static stack of layers
but rather as intermittent domains that are dynamically coupled together.  One
example is magnetic flux structures (or ``flux tubes'') fanning out with a
wine-glass geometry.  Such regular building blocks put certain constraints on
the implied atmospheric structure, which can make it difficult to fit in other
observational findings.  Accepting that magnetic field structures are far less
regular offers room for a more generally valid comprehensive picture.
This trend became more and more obvious during the recent years, both from
the observational and theoretical side \citep[see, e.g.,][and many
  more]{2007ASPC..368...49C, 2006ASPC..354..331G, 2007ASPC..369..193H,
  2006ASPC..354..259J, 2007ASPC..368...27R, 2007msfa.conf..321S}.

The advantages of a relaxed picture can be seen from the example of the quiet
Sun chromosphere above internetwork regions, which in itself is a complex and
intriguing phenomenon \citep[see, e.g.,][ and many
  more]{2006ASPC..354..259J,2006ASPC..354..276R,
  2004Natur.430..536D, 1999ApJ...517.1013L}.  Despite
tremendous progress, there are still many open questions concerning its
structure, dynamics and energy balance.  Recent observations now prove --
beyond any doubt -- the chromosphere to be a highly dynamic and intermittent
layer.  The internetwork chromosphere is the product of a dynamic interplay of
shock waves and magnetic fields.  This picture, which was already suggested by
many earlier investigations, offers a key to resolve some apparent
contradictions that lead to much confusion in the past.  A prominent example
concerns the observation of carbon monoxide \citep[see, e.g.,][]{ayres02},
which now can be explained as an integral part of a dynamic and intermittent
atmosphere
\citep{2007A&A...462L..31W,2006ASPC..354..301W,2005A&A...438.1043W}.  And
still the chromosphere cannot be investigated without also taking into 
account the layers above and
below.  The shock waves, which are so essential at least for the lowest,
weak-field parts of the chromosphere, are generated in the layers below, 
while  significant amounts of mass and energy are exchanged between the 
chromosphere and the corona above.
Obviously, the whole atmosphere must be seen as an integral phenomenon.

In the following sections, we report on a selection of results from
observations and numerical simulations, which will help us put together an
updated, revised view of the structure of the quiet Sun atmosphere.

\section{The Sun -- a multi-scale object}
\label{sec:aake}

An overarching point in this discussion is the fact that the Sun is
fundamentally a multi-scale object.  This is a major difficulty for
modeling and understanding, since it requires (computationally expensive)
modeling over a large range of scales.

\begin{figure}
\textbf{\large a)}\\[-5mm]
\centerline{\resizebox{0.8\hsize}{!}{\includegraphics*{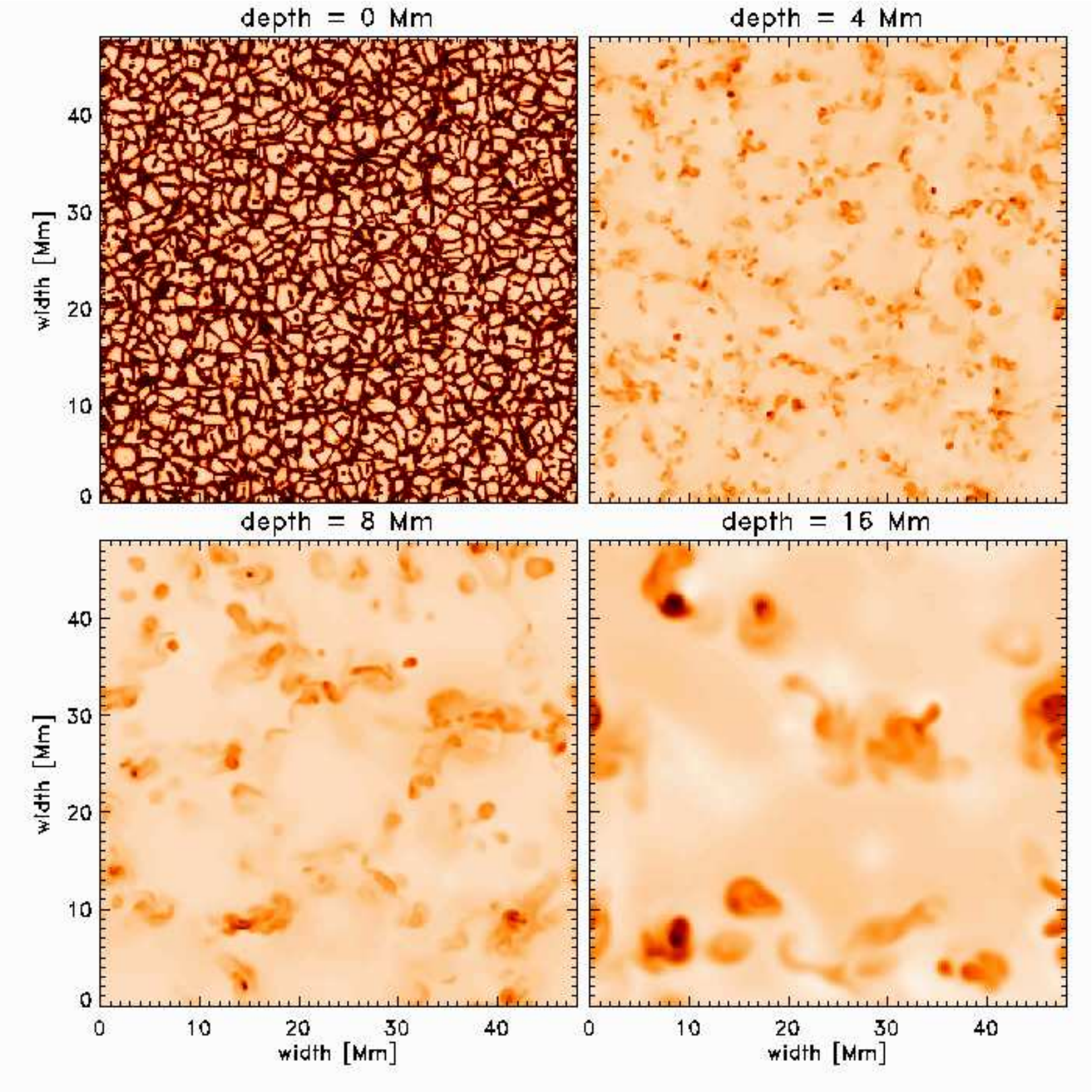}}}
{}
\textbf{\large b)}\\[-5mm]
\centerline{\resizebox{0.8\hsize}{!}{\includegraphics*{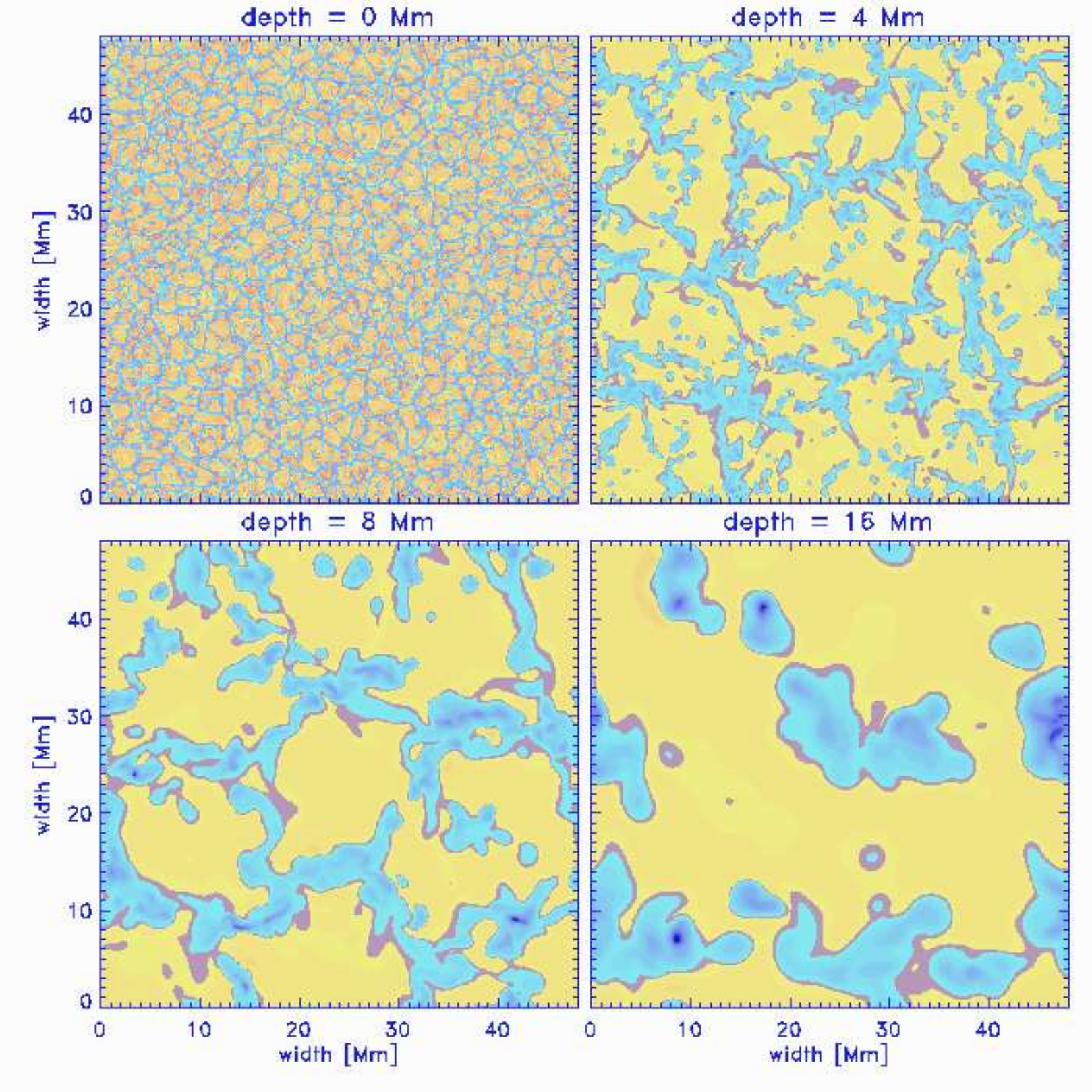}}}
\caption{Large scale solar convection (48\,Mm\,$\times$\,48\,Mm\,$\times$\,20\,Mm).
a)~Temperature (top four panels) and b)~velocity (bottom four panels) patterns
at four different depths.  
Temperature is shown on a linear scale.
Velocities are rendered with positive (downward) values blue and negative
(upward) velocities yellow.  A narrow band near zero velocity is rendered
in grey.
%
%
}
\label{fig:aake1}
\end{figure}
%
But the Sun also displays aspects of self-similarity and scale invariance in
several respects, which on the other hand helps a lot.  To illustrate the
self-similarity, Fig.~\ref{fig:aake1}a shows temperature patterns in horizontal
planes in a large scale simulation of solar convection
\citep{2007ApJ...659..848Z}, and Fig.~\ref{fig:aake1}b shows patterns of
vertical velocities from the same simulation.  The temperature patterns show
very intermittent cold structures, embedded in a background of horizontally
nearly constant temperature (images of entropy would look essentially
identical, with near-constancy also in the vertical direction).  
The set of panels also shows that the pattern scales increase systematically with depth.

Figure~\ref{fig:aake1}b, on the other hand, which displays vertical velocity on
a color scale 
that changes from yellow to blue with sign (with a narrow band
of grey for velocities near zero) gives a completely different impression.
With this rendering choice one can see that, at least from a morphological
point of view, the patterns at different depths are quite similar.  Displaying
in this way, signed velocity reveals that the sharply defined dark (cold)
patterns in Fig.~\ref{fig:aake1}a indeed correspond to the strongest downward
velocities, but that there are also relatively broad areas of much milder
downflows.  This shows that, as the ascending gas is forced (by mass
conservation) to overturn, it does so at first gently, then to finally be
accelerated more strongly by the positive feedback that comes from merging
with colder gas from above.

\begin{figure}
\centerline{\resizebox{\hsize}{!}{\includegraphics*{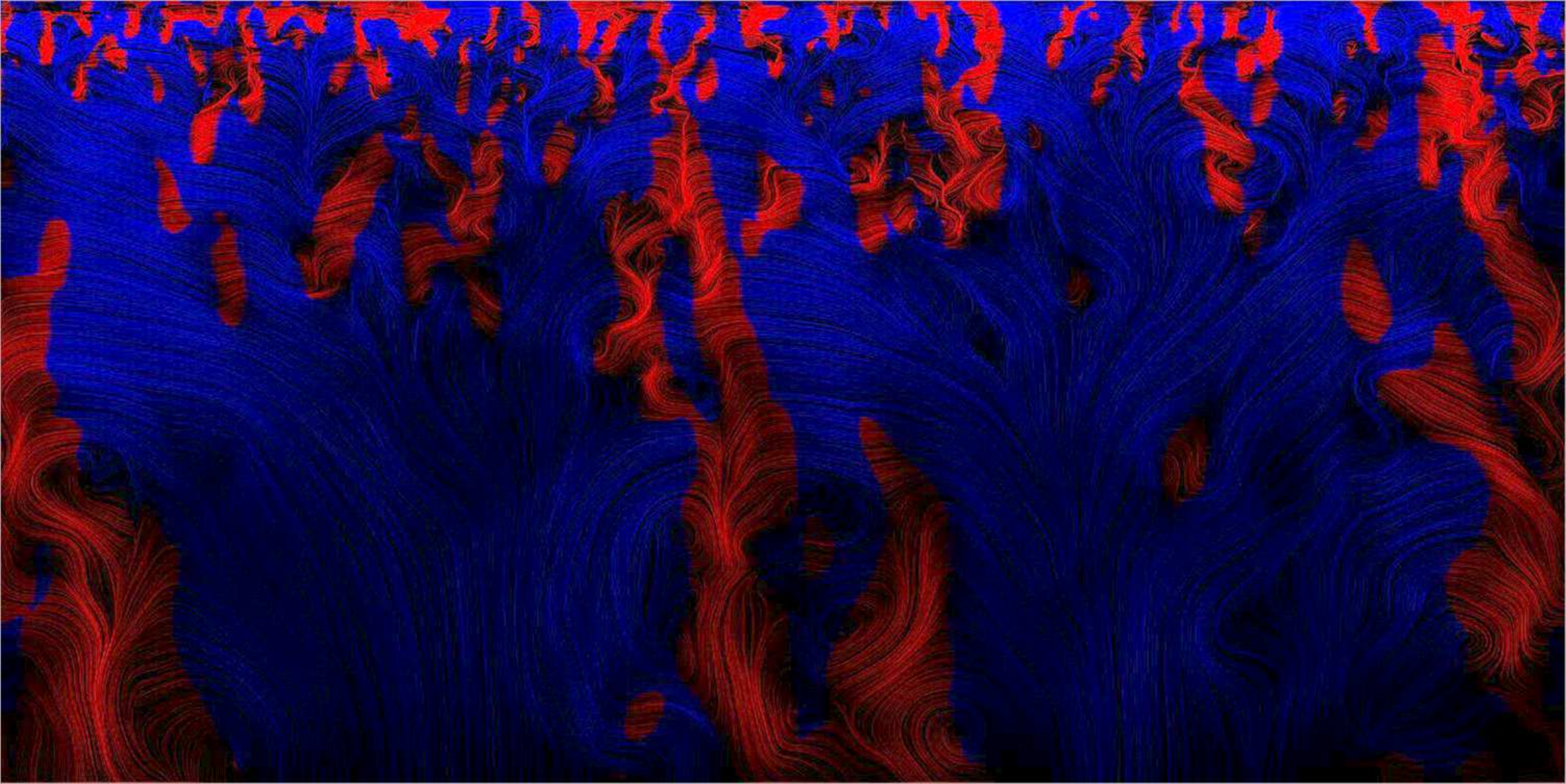}}}
\caption{A side view of a 48\,Mm\,$\times$\,48\,Mm\,$\times$\,20\,Mm simulation, 
showing velocity streamlines, with brightness increasing with increasing
magnitude.  
Up- and down-flows are rendered in blue and red, respectively.
}
\label{fig:aake7}
\end{figure}

At the visible surface the horizontal velocity patterns from various
depths are superposed.  This happens because the depth dependencies of the
large scale horizontal velocity patterns are rather weak; at least over
depth intervals small compared to their horizontal extents. On the other hand, as
illustrated by Fig.~\ref{fig:aake1}, the dominant scale becomes smaller
for layers increasingly close to the surface.  As the amplitudes of these smaller
scales are larger, they mask the presence of the larger scale patterns,
whose presence, however, can still be revealed, e.g., with Fourier analysis
or with low-pass filtering.  The hierarchy is illustrated in a side
view in Fig.~\ref{fig:aake7}.
The combination of streamlines and colors
illustrate how near-surface, small scales fluctuations are carried along
in larger scale flows.

\begin{figure}
\centerline{\resizebox{0.8\hsize}{!}{\includegraphics*{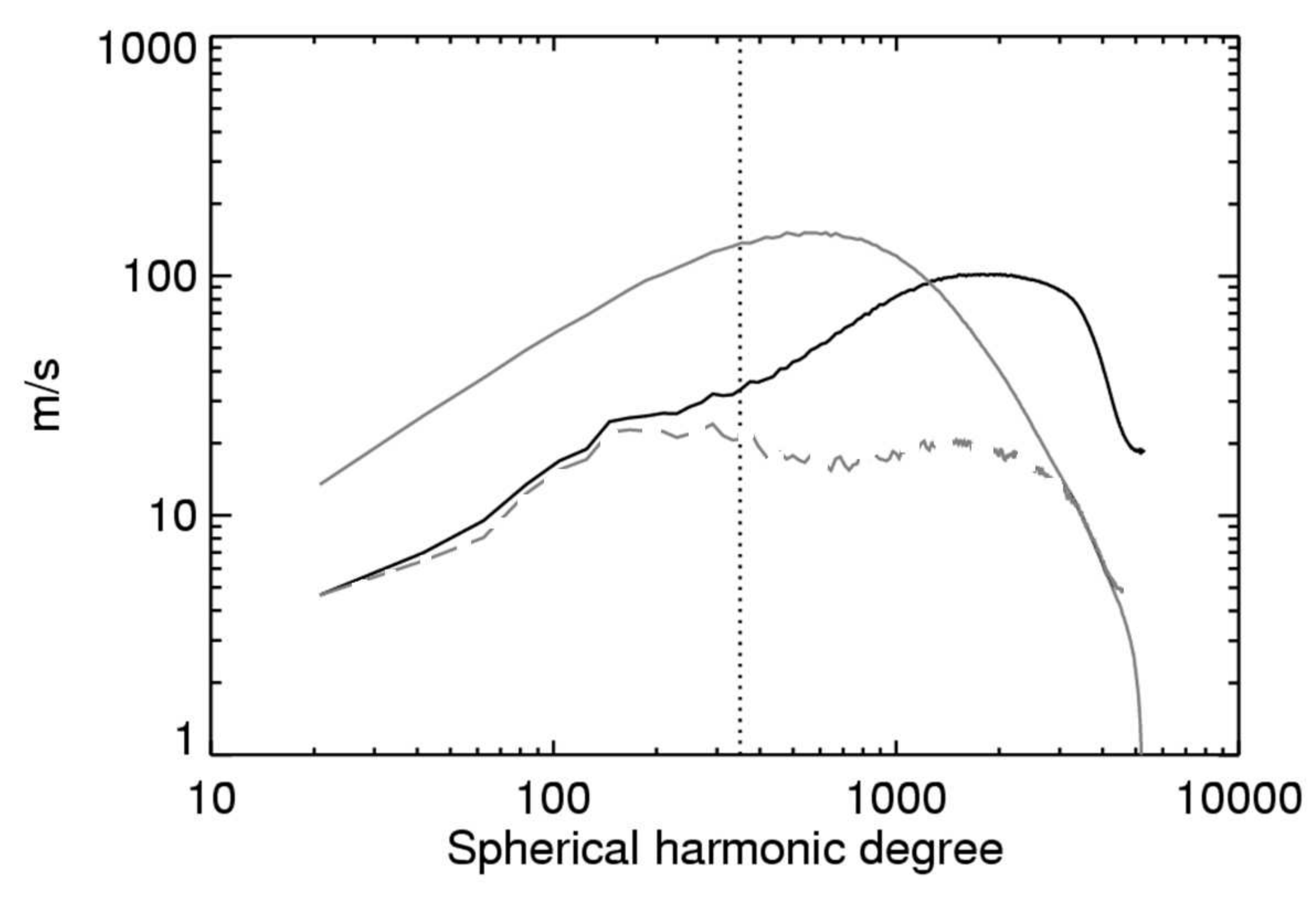}}}
\caption{Solar velocity spectrum:  A subsonic (7 km\,s$^{-1}$) filter has
been used to separate the velocity into oscillatory (grey) and
convective (black) components.  
The dashed line shows the convective
component resulting from first taking a 24h average.
Adapted from \citet{2007ApJ...657.1157G}.
}
\label{fig:aake3}
\end{figure}

The hierarchy of scales displayed reveals no particular preferred scale above
the granular one; the transition to larger and larger scales with depth is
smooth. A direct way to illustrate this from observations is to use power
spectra of solar velocities, as observed with SOHO/MID
\citep{2007ApJ...657.1157G}. Figure~\ref{fig:aake3} shows velocity (mainly
horizontal) as a function of size, produced by filtering the velocity 
power observed by MDI into sonic and sub-sonic parts.  The velocity spectrum
displayed is produced by then taking the square root of the velocity power
times wave number; this is a quantity --- a velocity spectrum --- that nicely
illustrates the dependence of velocity amplitude on size.  Note that there
is very little (less than a factor two) extra power at scales
traditionally associated with supergranulation, and that there is a smooth
and increasing distribution of velocity amplitude across the
``meso-granulation'' range to granulation scales.  The same behavior is
found in large scale numerical simulations \citep{2007ApJ...657.1157G}.

Supergranulation patterns can be brought forward by averaging over either
time or space; the dashed line in Fig.~\ref{fig:aake3} shows the effect of
a 24-hour time average.  A low-pass wavenumber spatial filter has a similar
effect; it cuts away the larger amplitudes at smaller scales and exposes aspects
of the underlying larger scale pattern. The relatively distinct appearance
of a supergranulation scale network in magnetically related diagnostics
indicates that the transport and diffusion of magnetic field structures
at the solar surface results in what is effectively a low-pass wavenumber
filter.

\begin{figure}
\centerline{\resizebox{0.4\hsize}{!}{\includegraphics*{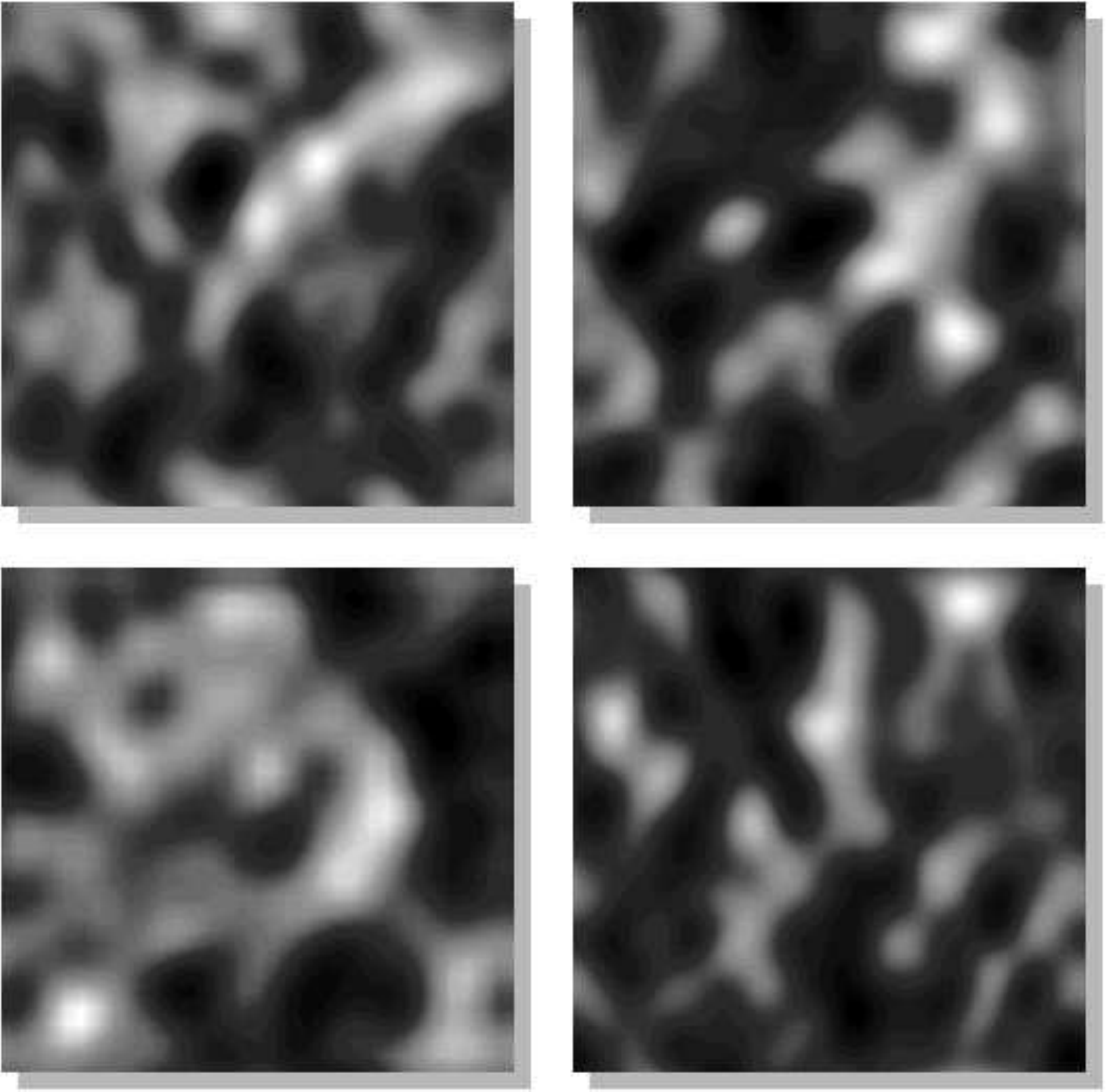}}}
\caption{Solar horizontal velocities observed with SOHO/MDI.  A patch some distance
away from solar center has been compensated for projection effects and
filtered to effective resolutions that differ by factors of 2. Which is
which?
}
\label{fig:aake4}
\end{figure}

As shown by Fig.~\ref{fig:aake4} it is practically impossible to tell the difference
between velocity patterns on different scales, once they are filtered to
effectively the same resolution.
As illustrated by \citet{2002ESASP.505..101S,2003AN....324..397S}
magnetic field patterns and distributions
also show a degree of self-similarity.

The magnetic energy equation
\begin{equation}
{\partial \over \partial t} \left(\frac{B^2}{8\pi}\right) =
    -  \nabla \cdot ({\bf E} \times {\bf B})
    - {\bf u} \cdot ({\bf j} \times {\bf B}) - Q_{\rm Joule}
\end{equation}
illustrates that balance of the magnetic energy at each depth is achieved by
Lorentz force work (by the flow on the field) being used to balance magnetic
dissipation, with net magnetic energy transported up or down by the Poynting
flux, ${\bf E}\times{\bf B}$.  As shown by the
work of \citet{2007A&A...465L..43V} 
and \citet{2008ApJ...680L..85S}
the actual direction of net transport is
systematically downwards, at least below the solar surface.  It appears likely
that there is net dynamo action at each depth in the convection zone, with net magnetic
energy delivered to the next layer down.  This naturally leads to a pile up
near the bottom of the convection zone.  The downward transport, which is
known from direct studies
\citep{1998ApJ...502L.177T,2001ApJ...549.1183T,2001A&A...365..562D}, is often
referred to as ``turbulent pumping'', and is associated with the asymmetry
between up and downflows (illustrated in Fig.~\ref{fig:aake1}b). 

On the largest scales (largest depths), and only there, differential rotation
enables a large scale global dynamo action, with patterns clearly controlled
by being stretched out by differential rotation. Buoyancy eventually pushes 
the fields back up.

Another evidence for self-similarity comes from the power law behavior of
flare energy distribution as a function of time.  This behavior is also
recovered in numerical simulations of 3D magnetic reconnection
\citep{1996JGR...10113445G}.  These are signs that magnetic reconnection occurs in a
multi-scale hierarchy, where  magnetic dissipation at large magnetic
Reynolds number (low
resistivity) creates a hierarchy. Large scale structures generate subsidiary
small scale structures, which do it again (on shorter time scales) and again,
until the spatial scales are small enough to support the dissipation.

Note the remarkable and wonderful argument, made already by Parker a long
time ago, which shows that driven magnetic dissipation must, if anything,
{\em increase} with decreasing resistivity---quite contrary to naive
expectations.
This has been verified in numerical experiments by at least three different
groups \citep{1996JGR...10113445G,1996ApJ...470.1192H,1999ApJ...527L..63D}.

The chromosphere and corona are likely to be heated in much the same way,
as is illustrated by the well known flux-flux relations between coronal and
chromospheric diagnostic. It is hard to even avoid, as in models of coronal heating 
there is a tendency of dumping much more energy in the chromosphere as a side effect
\citep{2002ApJ...572L.113G,2005ApJ...618.1020G}.

As pointed out by Phil Judge: The chromosphere is not a mess; the upper
chromosphere looks nearly force-free like the corona, whereas the lower
chromosphere is less force-free. Complexity comes from both the temperature
and density.  A central question is: What drives the flows (particularly the
cool upflows)?

\begin{figure}
\centerline{\resizebox{\hsize}{!}{\includegraphics*{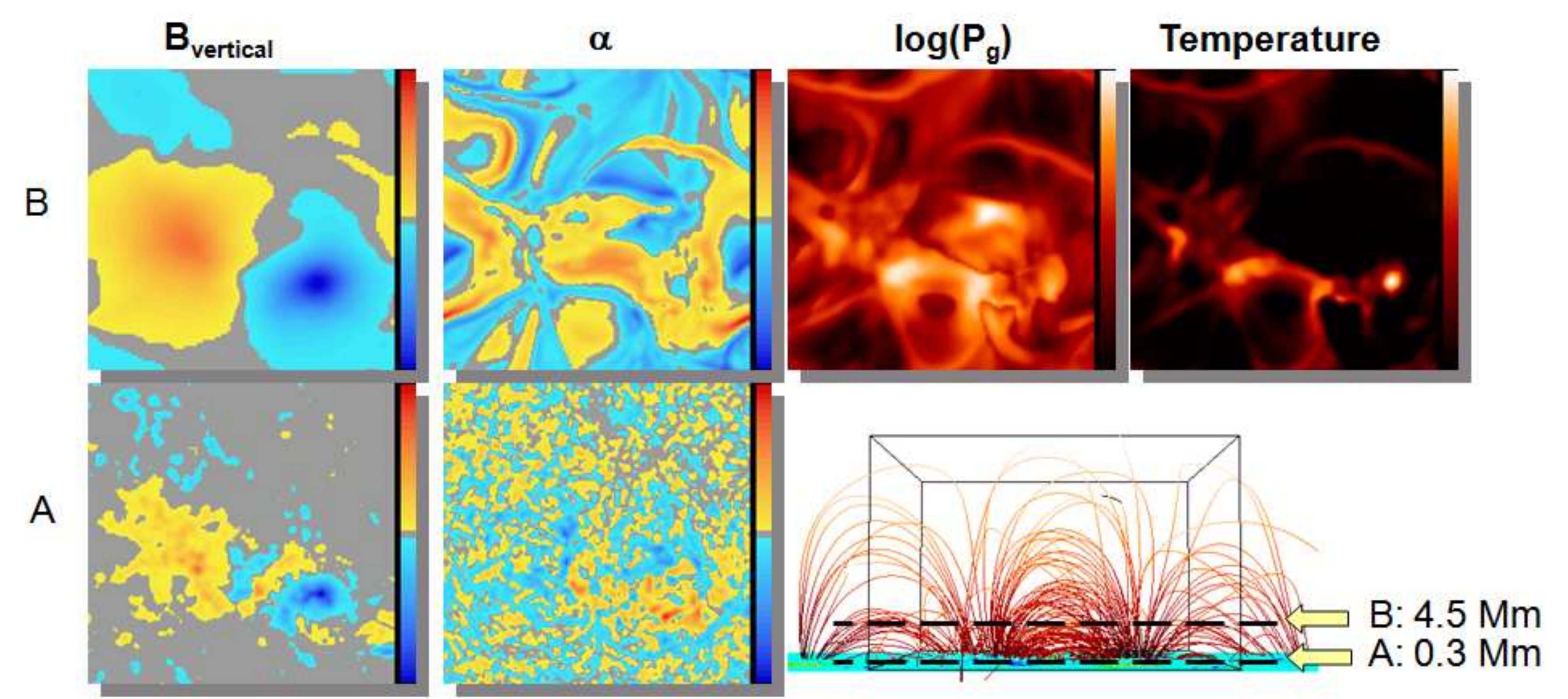}}}
\caption{The panels show the vertical magnetic field (far left), the ratio (often referred to with
the symbol $\alpha$) of the magnitude of the electric current along the
magnetic field to the magnitude of the magnetic field itself (left), the gas
pressure (right) and the log. temperature (far right).  The positions of
the cutting planes are indicated in the inset at bottom right.
Adapted from \citet{2005ApJ...618.1020G}.
}
\label{fig:aake5}
\end{figure}

Semi-realistic models of coupling of the horizontal photospheric velocity
field to the corona were first computed by \citet{2005ApJ...618.1020G}, who 
showed that a
correctly normalized photospheric (model) velocity field injects sufficient
power into the corona to create and maintain coronal temperatures (cf.\
Fig.~\ref{fig:aake5}).  

%
The mechanism is, in the absence of explicit flux emergence, essentially the
`braiding mechanism' introduced by
\citet{1972ApJ...174..499P,1981ApJ...244..631P,1983ApJ...264..642P}.
The heating is quite intermittent and drives up- and down-flows along
magnetic field lines into and out of the corona.
%
\begin{figure}
\centerline{\resizebox{\hsize}{!}{\includegraphics*{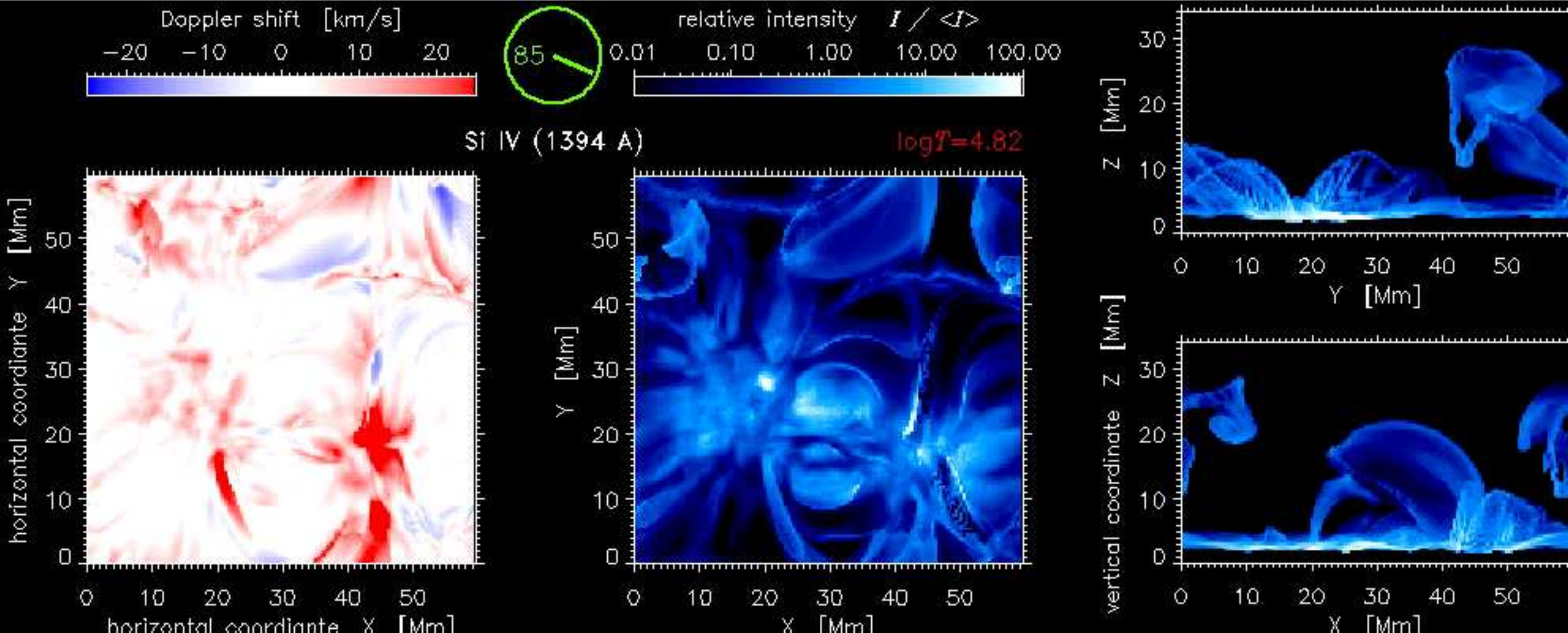}}}
\caption{The Si IV 1394 $\AA$ Doppler shift (left) and emission measure (mid) 
as they would be observed from above, and the emission measure projected along 
the Y-axis (right). 
Positive (downward) Doppler shifts are in red and negative 
Doppler shifts are in blue. Adapted from 
\edt{\cite{2004ApJ...617L..85P}.}
}
\label{fig:aake6}
\end{figure}
%
Emulated TRACE images and animations show qualitative agreement with
observations \citep{2004ApJ...617L..85P}.
Silicon IV, for example, picks up cooling condensations (cf.\ Fig.~\ref{fig:aake6}).
In addition, spectral lines formed at different temperatures show
semi-quantitative agreement of the dependency with depth of their
Doppler shifts and mean emission measures
\citep[cf.\ Figs.\ 7 and 9, ][]{2006ApJ...638.1086P}.
The differential
emission measure is a `fingerprint' type diagnostic, in much the same way as
spectral line asymmetries are for photospheric spectral lines
\citep{1981A&A....96..345D, 1990A&A...228..184D,2000A&A...359..669A,2000A&A...359..729A}.
Subsequently there has been much progress due to the work of the Oslo group
\citep{2005ESASP.592E..87H,2006ApJ...647L..73H,2007ASPC..368..107H,2008ApJ...679..871M}
-- cf.\ also the discussion in Sect.~\ref{sec:swb_lss}.

\section{Observations - Measuring the magnetic field in the solar atmosphere}
\label{sec:observ}

\begin{figure*}[p]
\resizebox{\hsize}{!}{\includegraphics*{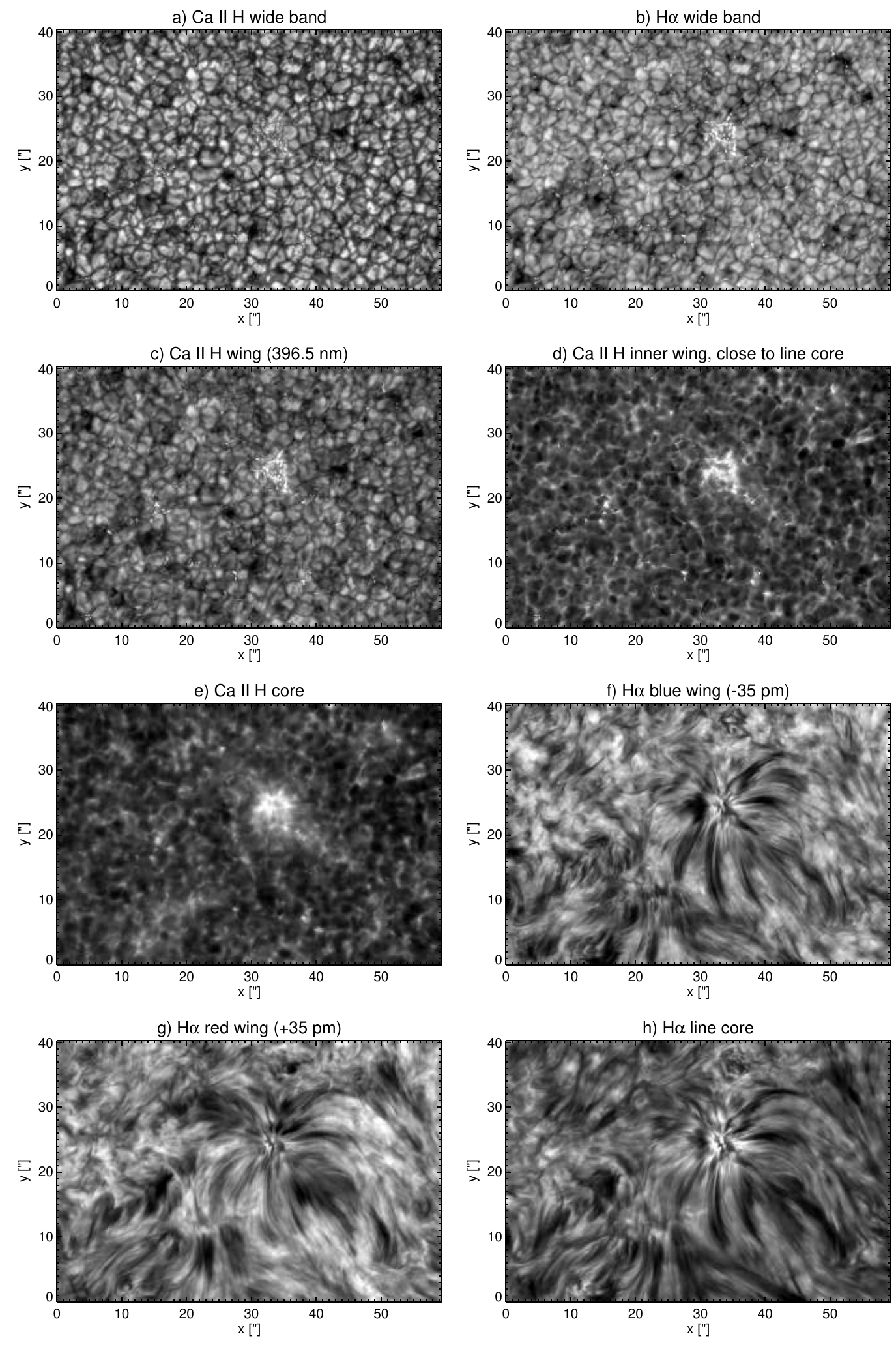}
}
\caption{Observations of a quiet Sun region close to disc-centre:
\textbf{a)}~Ca~II~H wide band,
\textbf{b)}~H$\alpha$ wide band image (FWHM 0.8\,nm),
\textbf{c)}~Ca~II~H wing (396.5\,nm),
\textbf{d)}~Ca~II~H inner wing  (close to line core),
\textbf{e)}~Ca~II~H core,
\textbf{f)}~H$\alpha$ blue wing at -35\,pm,
\textbf{g)}~H$\alpha$ red wing +35\,pm,
\textbf{h)}~H$\alpha$ line core.
The observations were carried out with the Swedish 1-m Solar Telescope (SST).
Data courtesy: L. Rouppe van der Voort (University of Oslo).
}
\label{fig:wln_obs8}
\end{figure*}

\begin{figure*}
  \centering
  \includegraphics[width=\linewidth,clip=TRUE]{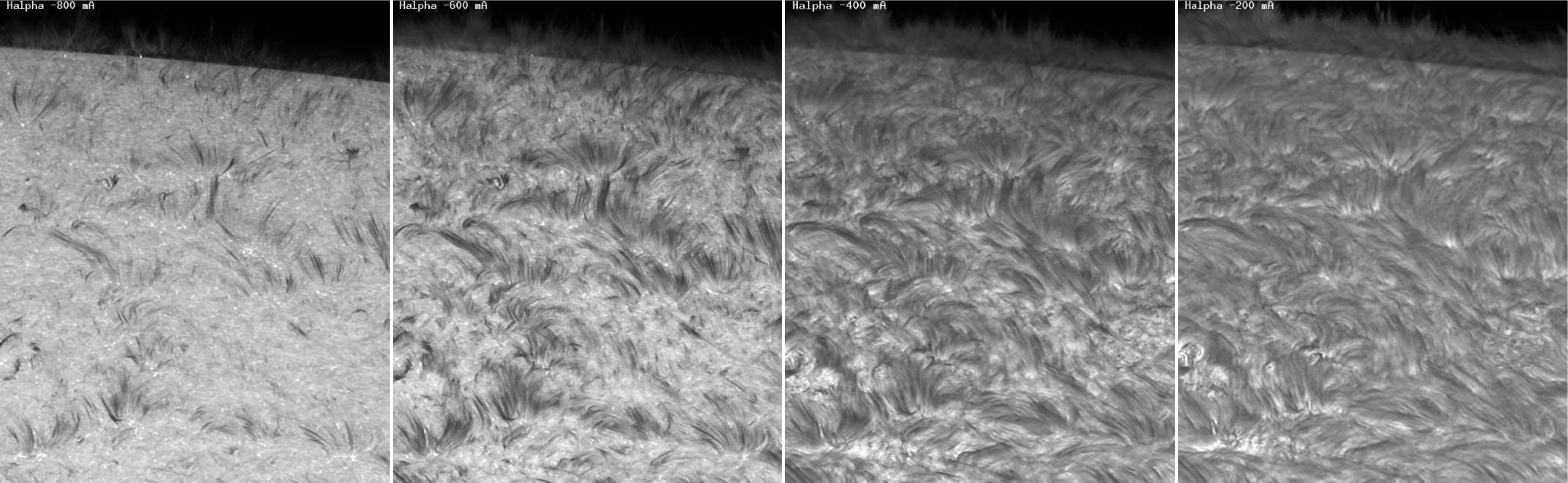}
  \caption{H$\alpha${} fine structure at -800, -600, -400, -200\,m\AA{} from
    line center recorded with the Dutch Open Telescope (DOT) on October 4,
    2005 \cite[taken from][]{2007ASPC..368...27R}. Cell-spanning fibrils are
    visible around line center (right). The decreasing line opacity in the
    blue wing of the line opens the view to the solar photosphere, intercepted
    by dark fibrils resulting from the Doppler shift of the line core.  }
  \label{halpha.pdf}
\end{figure*}

The H$\alpha${} line core images in Figs.~\ref{fig:wln_obs8}h and
\ref{halpha.pdf} show a well-known but still barely understood and intricate
picture: fibrils that spread from regions of enhanced magnetic field strength,
occasionally connecting to neighboring regions or apparently fading in between
\cite[see e.g.][]{2007ASPC..368...27R}. The structure gradually changes as one
goes from line center into the wings, as the corresponding intensity is due to
lower layers. Finally, a mesh-like background pattern shines through in the
internetwork regions. It is most likely due to reversed granulation in the
middle photosphere with some possible contributions from the low
chromosphere. The gradual change of the pattern in H$\alpha${} with wavelength
gives some clues about the atmospheric structure, in particular the magnetic
field in the chromosphere (the``canopy'' field), and definitely shows us that
the photosphere and chromosphere are coupled via magnetic fields on
medium to large spatial scales and via fields and shock waves on the small
scales.

Therefore, the understanding of the coupling between photosphere and corona is
intimately connected to the measurement of the chromospheric magnetic
field. The following subsections exemplify the difficulties of chromospheric
magnetic field measurements and present promising approaches to determine the
vector magnetic field of the chromosphere.

\subsection{Improving magnetic field extrapolations}

Measuring the magnetic field in the photosphere has a long tradition
\cite{hale:1908}. After 100 years of solar magnetic field measurements, the
level of sophistication, both in terms of instrumentation and in analysis
technique, has reached a very high level of maturity. Numerous telescopes,
ground based and space born  (e.g. GONG, MDI), investigate the global structure of the solar
magnetic field on a routine basis. High resolution
measurements allow the characterization of magnetic elements with a size in
the 100\,km range (e.g., SST - CRISP, Hinode SP). With HMI on SDO and the
balloon-borne 1\,m telescope Sunrise (launch: summer 2009) a major improvement
in the determination of photospheric magnetic fields will be achieved in both
directions - the global magnetic field configuration as well as the smallest
scale structures down to a size of 25\,km.

The availability of such high quality photospheric vector magnetograms and the
low plasma-$\beta$ in the chromosphere are the basic ingredients needed for
reliable, force-free magnetic field extrapolations. Starting with
\cite{sakurai:81} these extrapolations nowadays have reached a high level of
sophistication \cite[see reviews by][]{sakurai:89,amari:97,wiegelmann:08b}. To
further improve the accuracy of the chromospheric magnetic field
extrapolations additional information on the complex structure of the
chromosphere must be taken into account. One of the most promising advances in
this direction was proposed by \cite{wiegelmann:08a}: the basic assumption for
applying non-linear, force-free magnetic field extrapolations is the
force-freeness of the photospheric vector magnetograms. Measured magnetograms
do not fulfill this requirement, therefore a preprocessing of the measured
data is required. \cite{wiegelmann:08a} developed a minimization procedure
that yields a more chromosphere-like field by including the field direction
information contained in, e.g., chromospheric H$\alpha${} images. Including
this information into the extrapolation algorithm significantly enhances the
reliability of the extrapolations.

\subsection{Direct measurements of the chromospheric magnetic fields}

Measurement techniques for chromospheric magnetic fields have to overcome a
variety of hurdles: (i)~the plasma density is several orders of magnitude
lower than in the photosphere, (ii)~the energy transport is dominated by
radiation,
(iii)~the magnetic field strength is on average lower than in the photosphere,
and (iv)~anisotropic illumination induces population imbalances between atomic
sublevels that are modified by weak magnetic fields.  The low plasma density
leads to weak signals in the absorption (on-disk observations) or emission
(off-limb observations) of spectral lines. The absorption signatures of
chromospheric lines often show a strong photospheric contribution.  Only
highly spectrally resolved observations of the line core carry the
chromospheric information.  As a consequence of the low density, the
simplifying assumption of local thermodynamic equilibrium breaks down.  The
interpretation of the observations is thus by far more involved than in
the case of photospheric observations.
Additionally, the low chromospheric magnetic field strengths weakens the
Zeeman signals in spectral lines. Scattering polarization and its modification
by the Hanle effect introduce an additional complication in the analysis of
the polarization signal of spectral lines.

During the last decade major progress has been achieved in circumventing these
hurdles. Radio observations are able to determine the magnetic field strength
in and around active regions \cite[see review by][]{lee:07}.  Acoustic mapping
techniques \cite[]{finsterle:04} use the reflection of high-frequency acoustic
waves (mHz-range) from the region in the atmosphere where the gas pressure and
the magnetic pressure are equal to reveal the structure of the magnetic
canopy. The biggest leap in the direct determination of chromospheric magnetic
fields was achieved by combining state of the art instrumentation for full
Stokes polarimetry with recent progress in atomic physics. \cite{bommier:80},
\cite{landi:82}, \cite{stenflo:97c} and \cite{trujillobueno:02} opened a new
diagnostic window in solar physics: magnetic fields influence the strength and
the direction of the linear polarization resulting from atomic or scattering
polarization. This effect, discovered by \cite{hanle:24}, allows the
determination of the magnetic vector from Milligauss to several tens of Gauss,
a range not accessible by Zeeman diagnostics.

The following sections describe examples of measurements in this new
diagnostic window, focused around two of the most popular spectral lines for
combined Hanle and Zeeman measurements: the triplet of He~\textsc{i}
10830\,\AA{} and the He~\textsc{i}~D$_3$ 5876\,\AA{} multiplet.  The formation
of these lines requires ionization of para-He by ultraviolet radiation or
collisions, followed by recombination to populate the lower sates of
ortho-He. Since the main source for the ultraviolet radiation is the corona,
these He~\textsc{i} lines lacks almost any photospheric
contribution. Additionally, they are (generally) optically thin and narrow,
allowing the use of rather simple analysis techniques, like Milne-Eddington
inversions of the radiative transfer equations
\cite[]{solanki:03,lagg:04a,lagg:07a}.  With the inversion code HAZEL
\cite[HAnle and ZEeman Light, see][]{asensioramos:08a}, involving the joint
action of atomic level polarization and the Hanle and Zeeman effect in these
lines, a standard tool for the analysis of Stokes spectra is now available.

\subsubsection{Spicules}

Spicules are an ubiquitous phenomenon on the Sun.  At any time, the number of
these needle-like structures on the Sun is on the order of
4$\,\times$\,10$^5$. These dynamic and short lived features (lifetimes typically
$5-10$ minutes) can be considered as magnetic tunnels through which the
refueling of the coronal plasma takes place \citep{athay:00}.  High cadence
Hinode SOT observations in Ca~\textsc{ii}~H \cite[]{okamoto:07,suematsu:07}
revealed details in terms of size and dynamics and led to the discovery of a
new type of spicules \cite[type II spicules, ][]{2007PASJ...59S.655D} with
shorter lifetimes (10-150\,s), 
smaller diameters ($<$200\,km compared to
$<$500\,km for type I spicules), 
\edt{ and faster rise times. 
According to \citet[]{depontieu:07b}, they (i)~act as tracers for Alfv\'en waves 
with amplitudes of the order of 10 to 25\,km\,s$^{-1}$ 
and (ii)~carry, in principle, enough energy to 
play an important role for heating the quiet Sun corona and to 
accelerate the solar wind. 
}
See also Sect.~\ref{sec:canopydomain}.

\begin{figure}
  \centering
  \includegraphics[width=\linewidth,clip=TRUE]{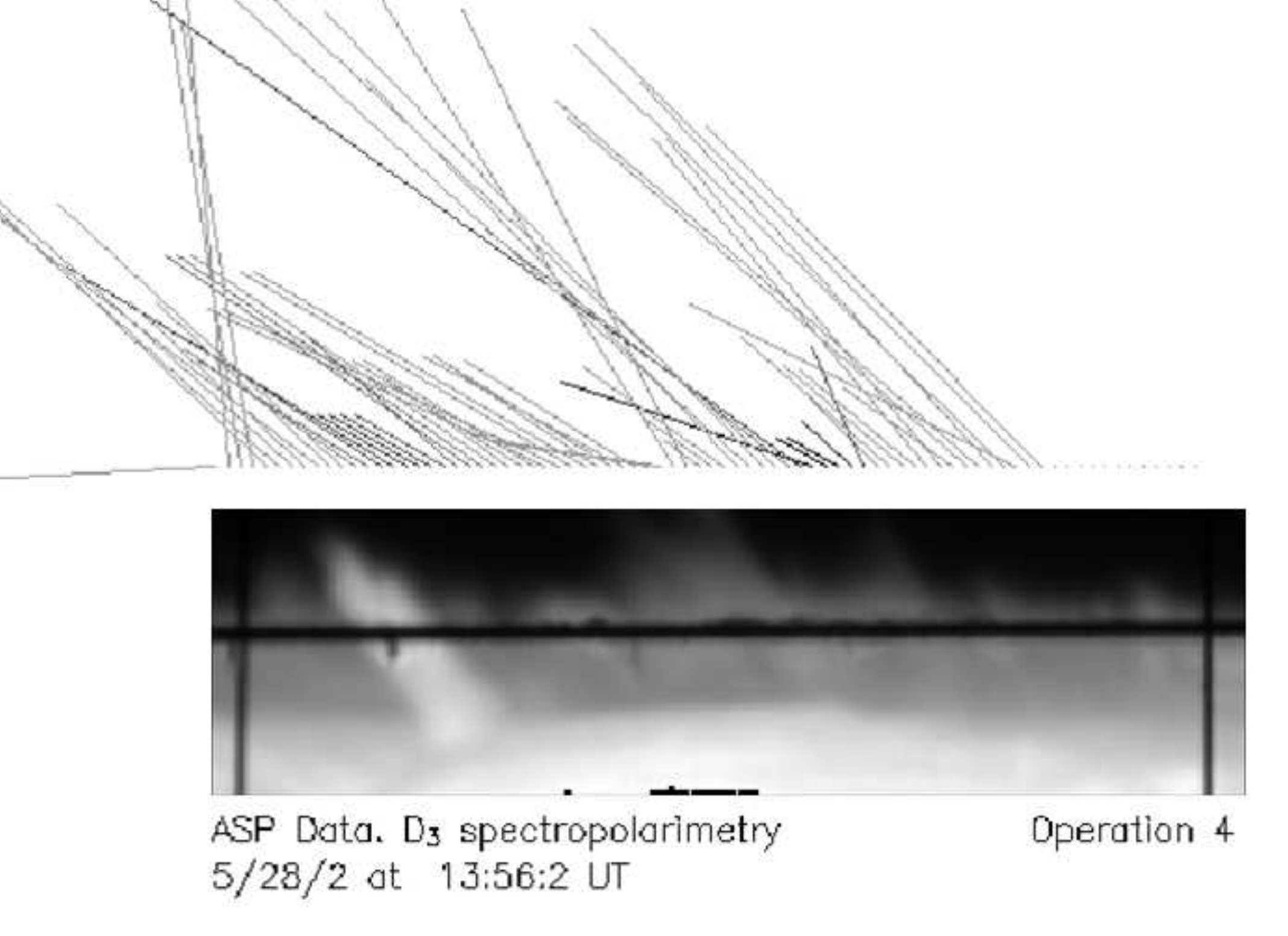}
  \caption{Magnetic field measurement in the He~\textsc{i} D$_3$ line: the
    magnetic field vector determined by a combined Hanle and Zeeman diagnostic
    traces the visible structures in the H$\alpha${} slit-jaw image
    \cite[bottom, adapted from][]{lopezariste:05}.}
    \label{lopezariste_spicule.pdf}
\end{figure}

Measurements of the magnetic field of spicules, both type I and type II, are
essential for the understanding of this phenomenon. \cite{trujillobueno:05}
were the first to directly demonstrate the existence of magnetized, spicular
material. Full Stokes polarimetric data in the He~\textsc{i} 10830\,\AA{}
line, obtained with the Tenerife Infrared Polarimeter \cite[]{collados:99},
were analyzed by solving the radiative transfer equation assuming an optically
thick atmosphere. The application of a combined Hanle and Zeeman diagnostic
revealed a magnetic field strength for the observed type I spicule of 10\,G
and an inclination angle of 37\mbox{$^{\circ}$} at a height of 2000\,km above
the photosphere. The authors state that 10\,G is the typical field strengths
for spicules at this height, but significantly stronger fields may also be
present. This result agrees with the measurements from \cite{lopezariste:05}
using full Stokes polarimetry in the He~\textsc{i} D$_3$ line. They find field
strengths not higher than 40\,G and a good correlation between the magnetic
field orientation and the visible structure in H$\alpha${} (see
Fig.~\ref{lopezariste_spicule.pdf}). An independent confirmation of these
measurements was presented by \cite{socasnavarro:05c} by using full Stokes
observations from SPINOR (Spectro-Polarimeter for INfrared and Optical
Regions, at the Dunn Solar Telescope). Their multi-line approach removes the
dependence of the strength of the Hanle signals on the zero-field polarization
produced by the scattering of anisotropic radiation in the higher atmosphere.

\subsubsection{Prominences and filaments}

\begin{figure*}
  \centering
  \includegraphics[width=\linewidth,clip=TRUE]{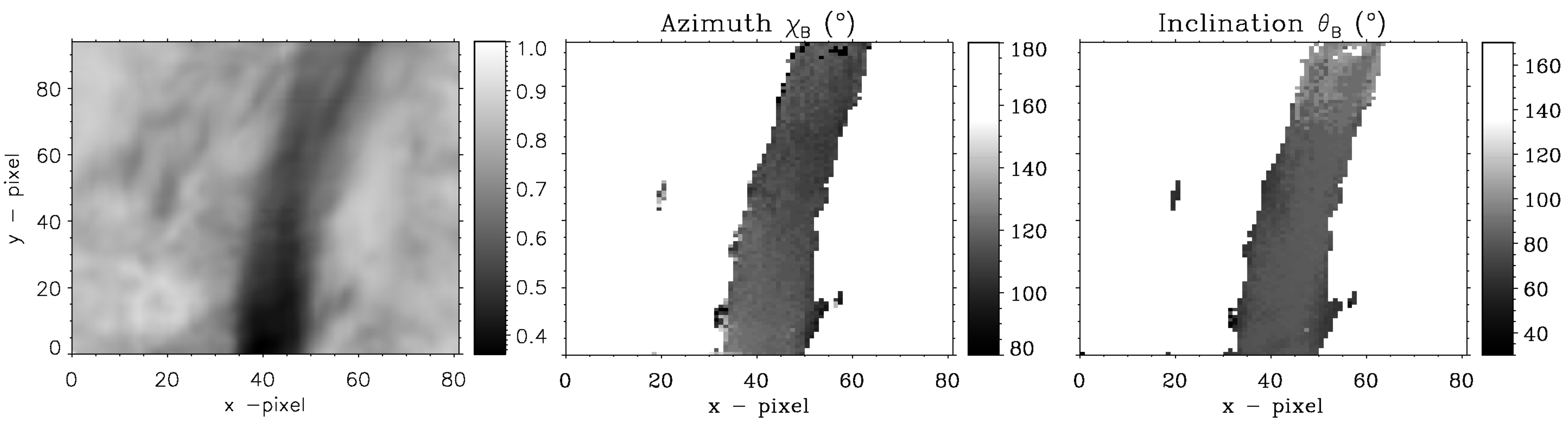}
  \caption{Intensity image in the center of the He~\textsc{i} 10830\,\AA{}
    line and derived azimuthal and inclination angle of the magnetic field
    \cite[adapted from][]{merenda:07a}.}
  \label{prominence.pdf}
\end{figure*}

The spectacular eruptions of prominences and filaments and the resulting
coronal mass ejections (CME) can cause sudden changes in the terrestrial
magnetosphere. A typical CME releases an energy of 10$^{25}$\,J and
10$^{12}$\,kg of solar material into the interplanetary space
\cite[]{harrison:94}. Before eruption, solar magnetic field holds this dense
and relatively cool material in the hot coronal environment and supports it
against the solar gravity for time periods as long as weeks. The knowledge of
the magnetic field within these structures therefore is of great interest to
understand the mechanisms leading to a possible eruption.

\cite{casini:03} were the first to present magnetic maps of prominences using
full Stokes polarimetry in the He~\textsc{i}~D$_3$ line. Their results confirm
previous measurements of the average field in prominences, ranging between 10
and 20\,G and oriented horizontally with respect to the solar
surface. However, they also find the presence of organized structures in the
prominence plasma embedded in magnetic field that are significantly larger
than average (50\,G and higher). \cite{merenda:07a} extended this work to
include the forward scattering case and applied it to a filament located at
disk center and obtained the first magnetic maps of a filament. In this
preliminary work they restricted their analysis to the saturated Hanle effect
regime between 10 and 100\,G. Here the linear polarization is only sensitive
to the direction of the magnetic field and does not change with intensity
variations. The results are reliable maps for the azimuth and inclination
angle for the magnetic field (see Fig.~\ref{prominence.pdf}). In agreement
with \cite{casini:03} they find horizontal fields in the central part of the
filament and a change of the azimuth according to the orientation of the main
axis of the filament. 
In order to detect, for example, the small-scale and rapidly moving 
filaments mentioned in Sects.~\ref{sec:swb_inc} and \ref{sec:fluctosphere}, 
significant improvements in signal to noise ratio and temporal resolution 
of polarimetric observations in this spectral line are required. 
The complex magnetic field and velocity structure of an
erupting filament in the He~\textsc{i} 10830\,\AA{} line was analyzed by
\cite{sasso:07b}: besides the magnetic field topology they identify the
presence of up to 5 different atmospheric components, distinguished by their
velocities ranging from -50 to +100\,km\,s$^{-1}$, within the resolution
element of approximately 1.$^{\prime\prime}$5. This measurement clearly
demonstrates the fibrilar structure of the chromosphere \cite[see
  also][]{lagg:07a} and the need for higher spatial resolution measurements in
this line.

\begin{figure}
  \centering
  \includegraphics[width=\linewidth,clip=TRUE]{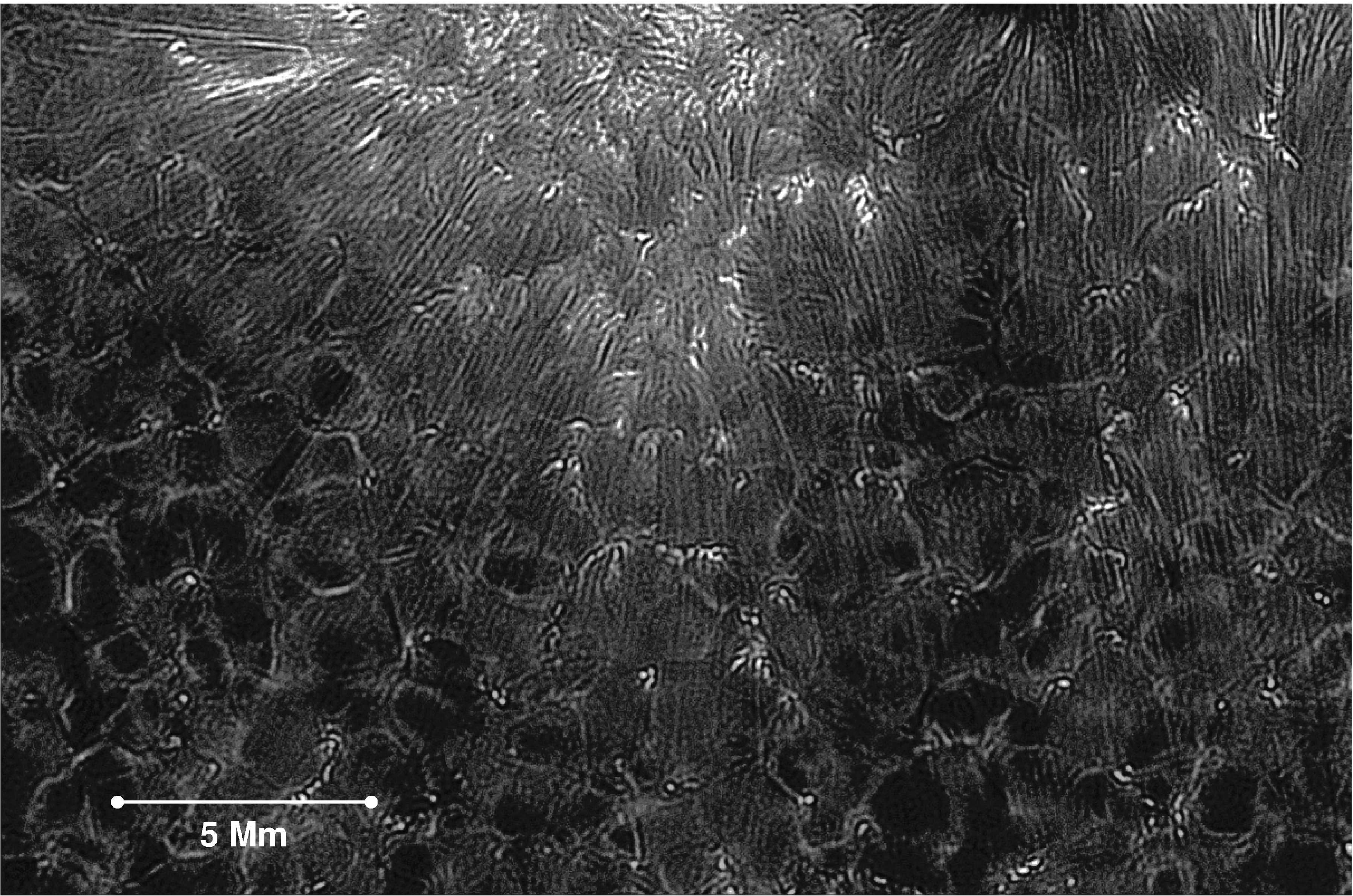}
  \caption{Speckle-reconstructed, narrow band image (contrast-enhanced) of a
    plage region observed in the line core of Ca~K using the SST
    \edt{\citep[][cf. \citeauthor{2005A&A...435..327R}\ \citeyear{2005A&A...435..327R}]{pietarila:08a}. }
    The Ca~K fibrils extend over quiet Sun regions. 
    The mesh-like background pattern is nevertheless dominated by the reversed 
    granulation pattern in the middle photosphere (cf. Fig.~\ref{fig:wln_obs8}).}
    \label{cak_fibrils_label.pdf}
\end{figure}

\subsubsection{Canopy}

Following previous work by W.~Livingston, \cite{gabriel:76} introduced the
term canopy to explain the emission measures of chromospheric and transition
region UV lines. In the ``classical'' picture, the magnetic pressure wins over
the gas pressure with increasing height, so that the magnetic flux
concentrations rooted in the network expand and cover the internetwork cells
with horizontal fields (see Sect.~\ref{sec:swb_atm} for an updated view).
\cite{giovanelli:82,jones:82} performed detailed studies of the magnetic
canopy close to the limb by determining magnetograms using chromospheric
spectral lines like the Ca~\textsc{ii} triplet at 8542\,\AA{} or the
Mg~\textsc{i}\,b$_2$ line at 5173\,\AA{}. These magnetograms are characterized
by a polarity inversion line parallel to the limb, on either side surrounded
by diffuse fields above the internetwork region \cite[see][for a sketch of the
  magnetic configuration]{steiner:00}.

Especially during the last decade diagnostic tools involving the Hanle effect
significantly improved the possibilities to characterize the canopy fields.
Using spectropolarimetric data in the Sr~\textsc{ii} 4078\,\AA{} line ``Hanle
histograms'', showing the statistical distributions of the Hanle rotation and
depolarization effects, \cite{bianda:98b} determined the magnetic field
strength of horizontal, canopy-like fields to be in the range of 5 to
10\,G. The first spatial mapping of Hanle and Zeeman \cite[]{stenflo:02a}
effect revealed details of canopy fields in a semi-quiet region measured close
to the limb in the Na~\textsc{i}~D$_1$-D$_2$ system. The authors found direct
evidence for horizontal magnetic fields, slightly stronger than the field
strengths determined by \cite{bianda:98b} (25-35\,G), that remain coherent
over a spatial scale of at least three supergranules.

\begin{figure}
  \centering
  \includegraphics[width=\linewidth,clip=TRUE]{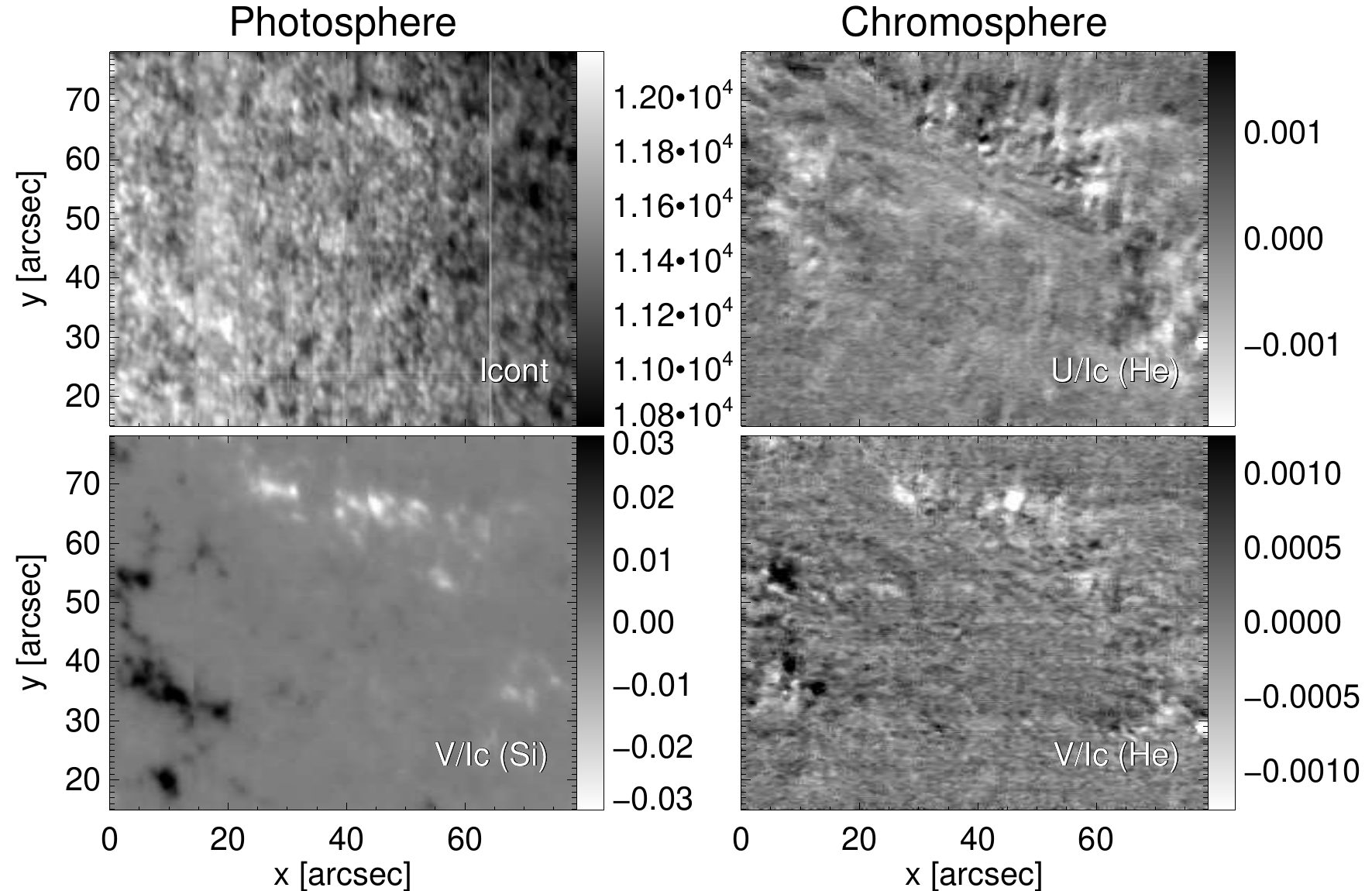}
  \caption{Measurement of the magnetic field over a supergranular cell in the
    photosphere and the chromosphere (German Vacuum Tower Telescope, Tenerife
    Infrared Polarimeter 2, May 10, 2008): continuum close to the 10830\,\AA{}
    line (top left), Stokes $V$ signal integrated over the red wing of the
    photospheric Si~\textsc{i} 10827\,\AA{} line (bottom left), Stokes $U$ and
    $V$ signal integrated over red wing of the chromospheric He~\textsc{i}
    10830\,\AA{} line (top and bottom right, respectively). The chromospheric
    maps suggest the presence of magnetic structures organized on mesogranular
    scales within the supergranular cell outlined by the photospheric Stokes
    $V$ map.}
    \label{canopy.pdf}
\end{figure}

The concept of a magnetic canopy around sunspots and in active regions is well
established. Over quiet regions, the formation of this layer of horizontal
fields is matter of debate: \cite{2003ApJ...597L.165S} showed that
concentrations of magnetic flux in the network in the order of a few tens of
Mx\,cm$^{-2}$ will destroy the classical, wineglass-shaped magnetic field
topology. Such flux concentrations, suggested by simulations, were identified
by \cite{2004Natur.430..326T} in terms of ubiquitous tangled magnetic field
with an average strength of $\approx$130\,G, much stronger in the
intergranular regions of solar surface convection than in the granular
regions. A significant fraction of this hidden magnetic flux has now been
clearly identified with the spectropolarimeter of the Hinode spacecraft
\cite[]{lites:08a}. However, narrow-band (0.1\,\AA{}) observations in the Ca~K
line with a spatial resolution of 0.$^{\prime\prime}$1 obtained with the
Swedish Solar Telescope (SST) provide evidence that magnetic fibrils,
originating from network flux concentrations, do span over a large distance
above the quiet Sun network 
\edt{(see Figs.~\ref{fig:wln_obs8} and \ref{cak_fibrils_label.pdf}).} 
Magnetic field
measurements using the He~\textsc{i} 10830\,\AA{} line also indicate the
presence of a uniform, horizontal magnetic field topology over the
internetwork at mesogranular scales \cite[]{lagg:08a}. These measurements,
presented in Fig.~\ref{canopy.pdf}, were obtained with the Tenerife Infrared
Polarimeter II (TIP-2) mounted behind the Vacuum Tower Telescope (VTT) on
Tenerife \cite[]{collados:07a} at a heliocentric angle of 49\mbox{$^{\circ}$}
($\mu=\cos\Theta=0.65$). Inversions involving Hanle and Zeeman effect prove
the presence of a horizontal ``canopy'' magnetic field on mesogranular scales
with strengths of the order of 50 to 100\,G, similar to the value of the
averaged magnetic field of the underlying photosphere. Both, \edt{recent} narrow-band
Ca~K observations of \citep[e.g.,][]{pietarila:08a,2005A&A...435..327R} and magnetic field measurements
\edt{
\citep[e.g.,][]{lagg:08a}} 
\edt{seem to be} 
in {\em apparent} contradiction to
\cite{2003ApJ...597L.165S}, pointing out the necessity of a more detailed
analysis on the validity of the concept of the magnetic canopy over quiet Sun
regions. Nevertheless, the different finding can be fit into a common picture,
when taking into account the sampled height ranges and a field topology, which
is more complex and entangled on small scales than usually assumed (see
Sect.~\ref{sec:swb_atm}).  The ``classical'' canopy might be in some ways a
too simplified and thus potentially misleading concept.

\section{Numerical simulations of the quiet Sun}


\subsection{Internetwork photosphere} 
\label{sec:swb_inp}

The solar granulation is now well reproduced by modern radiation
(magneto-)hydrodynamical simulations.  The contrast of continuum intensity
or ``granulation contrast'' is often used for comparisons between observations
and simulations.  
For many years, the contrast derived from observations were
much lower than those found in numerical simulations. 
One reason is the
often unknown but crucial effect of an optical instrument and the Earth
atmosphere, resulting in a significant decrease of the granulation contrast.
This problem can partially be overcome by using observations with space-borne
instruments.  Recent observations with the Broadband Filter Imager (BFI) of
the Solar Optical Telescope (SOT) onboard the Hinode spacecraft now show
higher contrast values.  After application of a realistic point spread
function 
\citep{2008A&A...487..399W,2008A&A...484L..17D},
state-of-the-art numerical simulations indeed reproduce important
characteristics of ``regular'' granulation.

\begin{figure}
\resizebox{\hsize}{!}{\includegraphics*{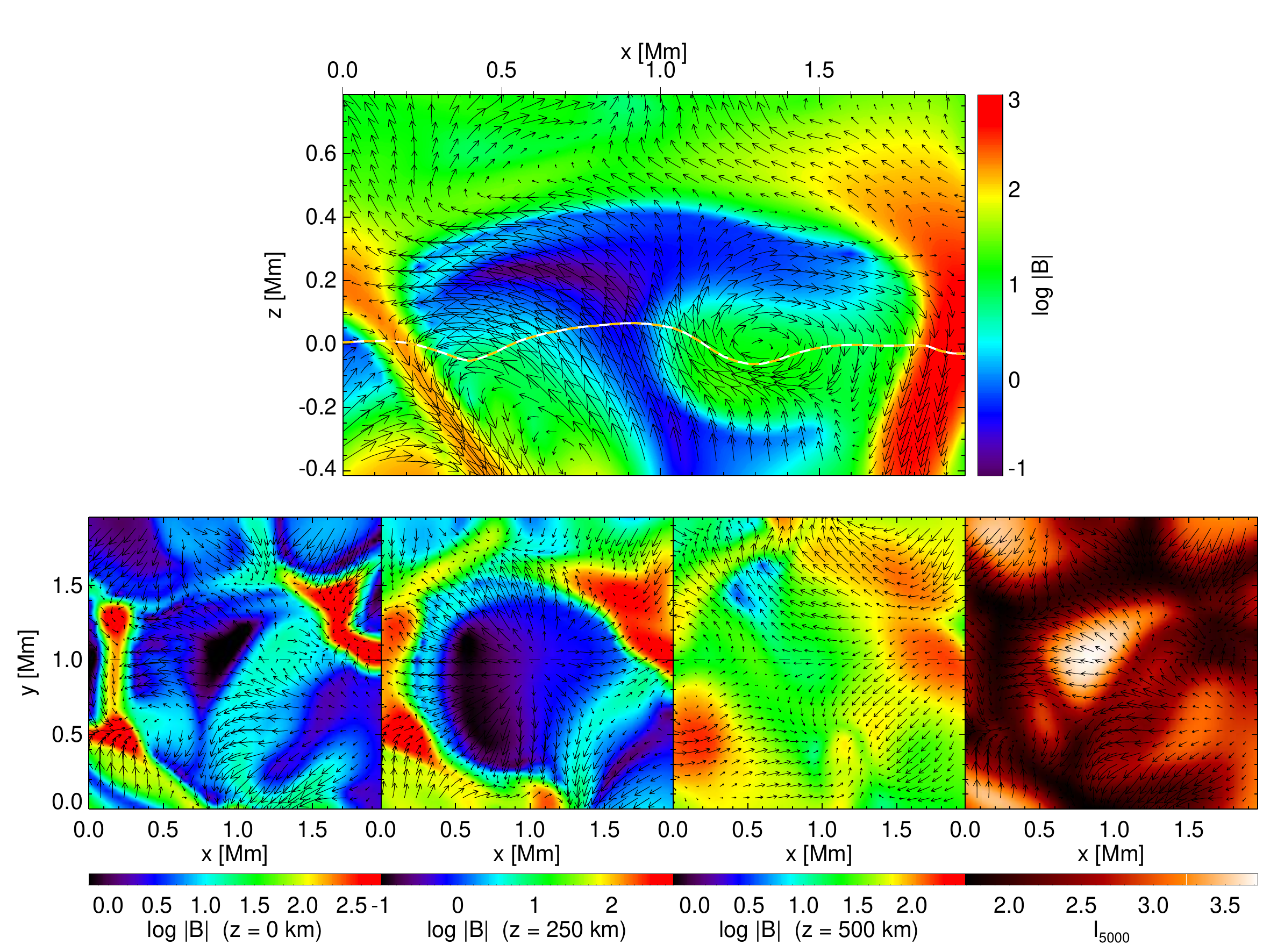}}
\caption{Flux expulsion in a close-up from a MHD simulations by
  \citet{2005ESASP.596E..65S}: Logarithmic magnetic field strength in a
  vertical cross-section (top) and in three horizontal cross-sections (bottom)
  at heights of 0\,km, 250\,km, and 500\,km.  The emergent intensity is
  displayed in the rightmost panel.  The arrows represent the velocity field
  in the shown projection planes.  The white line in the upper panel marks the
  height of optical depth unity.  }
\label{fig:swb_fluxexp}
\end{figure}

The convective flows in and just above granule interiors advect magnetic field
laterally towards the intergranular lanes, where the field is concentrated in
knots and sheets with up to kilo-Gauss field strengths.  In the granule interiors,
usually only weak field remains, although in some situations flux
concentrations of up to a few hundred Gauss can occur within the granules
\citep{2008ApJ...680L..85S}.  The latter finding is in agreement with the
observations by \citet{2007ApJ...666L.137C} and \citet{2008A&A...481L..25I}.
This process of ``flux expulsion'' has been known since early simulations
\citep{1981ApJ...243..945G,1986ssmf.conf...83N}.  It is now an integral part
of magnetoconvection simulations \citep[see,
  e.g.,][]{1996MNRAS.283.1153W,1998ApJ...499..914S,
  1998ApJ...495..468S,2005ESASP.596E..65S,voegler05}.  The close-up from a
simulation by \citet{2005ESASP.596E..65S} in Fig.~\ref{fig:swb_fluxexp}
illustrates the process.  The magnetic field in the low photosphere is not
only advected laterally.  It is also lifted upwards and is concentrated above
the reversed granulation layer at a height, which roughly corresponds to the
classical temperature minimum in semi-empirical models \citep{fal93}.  There,
the convective overshooting effectively dies out and most of the upward
directed flows above the granule interiors turn into lateral flows
(Fig.~\ref{fig:swb_fluxexp}.  In the models by \citep{wedemeyer04a,
wedemeyerphd}, the rms velocity amplitudes are smallest at these heights.
The result is that the magnetic field is ``parked'' there and forms a mostly
horizontally aligned field.  It connects to the photospheric flux funnels,
which spread out from the intergranular lanes below.  The enclosed regions
below, on the other hand, are virtually field-free with field strengths of
possibly down to a few Gauss only \citep{2003A&A...406.1083S}.  The field
configuration around these granular voids was referred to as a dynamic
``small-scale canopy'' by \citet{2005ESASP.596E..65S,2006ASPC..354..345S}.
Virtually field-free granule interiors are very common in their simulations.
In the more recent simulations by \citet{2008ApJ...680L..85S}, this phenomenon
is also existent but less pronounced, although the field in the granular
interiors is still much weaker than in the surrounding lanes.  The main
differences between these simulations is the average field strength (10\,G and
20\,G, resp.) and the injection of horizontal field at the lower boundary in
the latter simulation.  Obviously, the exact occurrence of small-scale
canopies still depends on details of the simulations and thus needs to be
checked by comparison with observations.  The recent detection of so-called
``horizontal inter-network fields'' (HIFs) can be regarded as observational
support for the small-scale field structure seen in the simulations.  It is
observed that the horizontal field component in the granular interiors is
stronger than the vertical component
\citep{1996ApJ...460.1019L,2007ApJ...670L..61O,
  2007PASJ...59S.571L,2008ApJ...672.1237L}.  HIFs are also clearly present in
simulations
\citep{2006ASPC..354..345S,2008A&A...481L...5S,2008ApJ...680L..85S} and are in
good agreement with the observations.

The direction of the horizontal magnetic field, which is continuously lifted
to the upper photosphere and lower chromosphere, varies.  Consequently,
current sheets form where different field directions come close to each
other.  In the simulations by \citet{2006ASPC..354..345S}, a complex stacked
meshwork of current sheets is generated at heights from $\sim 400$\,km to
$\sim 900$\,km.  The lower limit of this range, which is the typical height of
the small-scale canopies can be considered as the upper boundary of the
photosphere.

\subsection{Internetwork chromosphere} 
\label{sec:swb_inc}

\begin{figure} 
\resizebox{\hsize}{!}{\includegraphics*{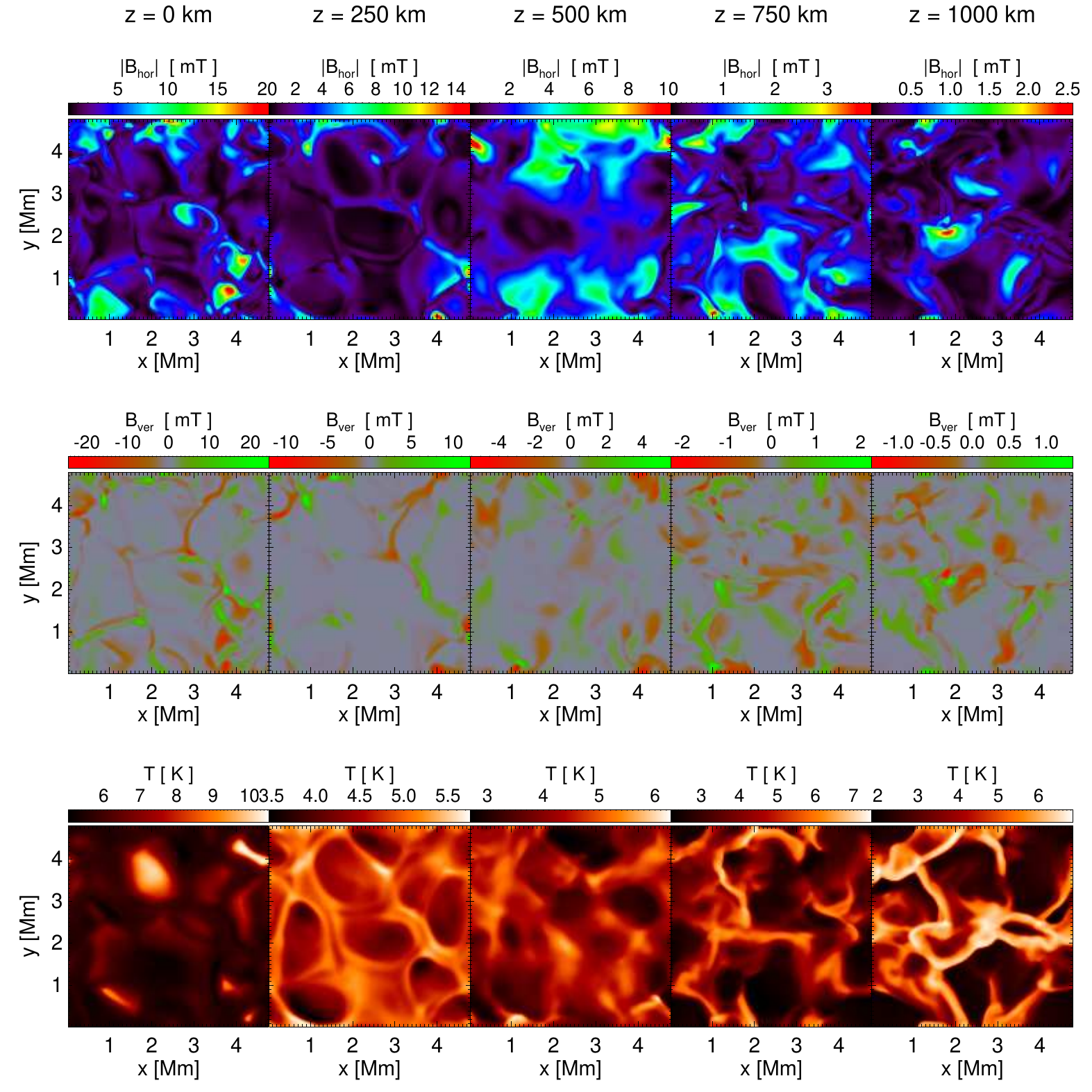}}
\caption{Horizontal cross-sections through the model by
  \citet{2005ESASP.596E..65S,2006ASPC..354..345S} showing the horizontal
  magnetic field component (top), the vertical component (middle row), and the
  gas temperature (bottom) at different heights: $z = 0$\,km (granulation),
  250\,km (reversed granulation), 500\,km, 750\,km (fluctosphere), and
  1000\,km (from left to right).  }
\label{fig:swb_xyslices}
\end{figure}

In recent years models have been extended in height to include the
chromosphere.  Modeling this layer is an intricate problem as many
simplifying assumptions, which work fine for the lower layers, are not valid
for the thinner chromosphere.  Rather, time-dependent three-dimensional
non-equilibrium modeling is mandatory.  This is in particular true for the
radiative transfer, for which deviations from the (local) thermodynamic
equilibrium should be taken into account.  Numerically, this is a demanding
task.  It is unavoidable to make simplifications and compromises when
implementing at least the most important non-equilibrium effects in a
time-dependent multi-dimensional simulation code.  A practicable way is to
start with simplified models and increase the amount and the accuracy of
physical ingredients step by step.  In their pioneering work,
\citet{carlsson94, carlsson95} implemented a detailed radiative transfer,
which was affordable by restricting the simulation to one spatial dimension.
This simplification made it necessary to implement an artificial piston below
the photosphere to excite waves as the convection cannot be realistically
simulated in one spatial dimension.  The high computational costs for such
detailed radiative transfer calculations forced \citet{skartlien00c} to use a
simplified description for their three-dimensional model.  Nevertheless, their
treatment included scattering.  Simplifications of the radiative transfer are
necessary for three-dimensional simulations in order to make them
computationally feasible.  This class of 2D/3D numerical simulations cover
a small part of the near-surface layers and extent vertically from the upper
convection zone to the middle chromosphere.  This way the shock-waves are
excited by the simulated convection without any need for an artificial driver.
The chromospheric layer of these models is usually characterized by intense
shock wave action, putting high demands on the stability of numerical codes.
\citet{wedemeyer04a} made experiments with simplified 3D models without
magnetic fields, using CO$^5$BOLD \citep{cobold}.  As in the aforementioned
simulations, they found that overshooting convection in the photosphere
triggers acoustic waves that propagate upwards and steepen into shock fronts.
The result is a dynamic layer above a height of $\sim 700$\,km, which is
composed of hot shock fronts and cool post-shock regions.  The gas temperature
in horizontal cross-sections through the model exhibits highly dynamic
mesh-like pattern with spatial scales comparable to the granulation.  The same
can be seen in the follow-up simulations by \citet{2005ESASP.596E..65S}, which
include weak magnetic fields (see Fig.~\ref{fig:swb_xyslices}).  The gas
temperature in the \mbox{CO$^5$BOLD} model chromospheres range from about
7000\,K down to 2000\,K, owing to the adiabatic expansion of the post-shock
regions.  A similar pattern is also present in the simulations by
\citet{2008ApJ...679..871M}.  The temperature range is very similar in both
models, but the temperature amplitudes differ.  Some snapshots of the
simulation by \citeauthor{2008ApJ...679..871M} also show a double-peaked
temperature distribution at chromospheric heights, but the cool background
component is usually much weaker than in the \mbox{CO$^5$BOLD} model.
Possible reasons for the differences are related to the numerical treatment of
the radiative transfer in the upper layers.  A shock-induced pattern can
already be perceived in the temperature maps by \citet{skartlien00c}, although
it less pronounced due to the relatively coarse grid spacing in this earlier
simulation.

\begin{figure}
\resizebox{\hsize}{!}{\includegraphics*{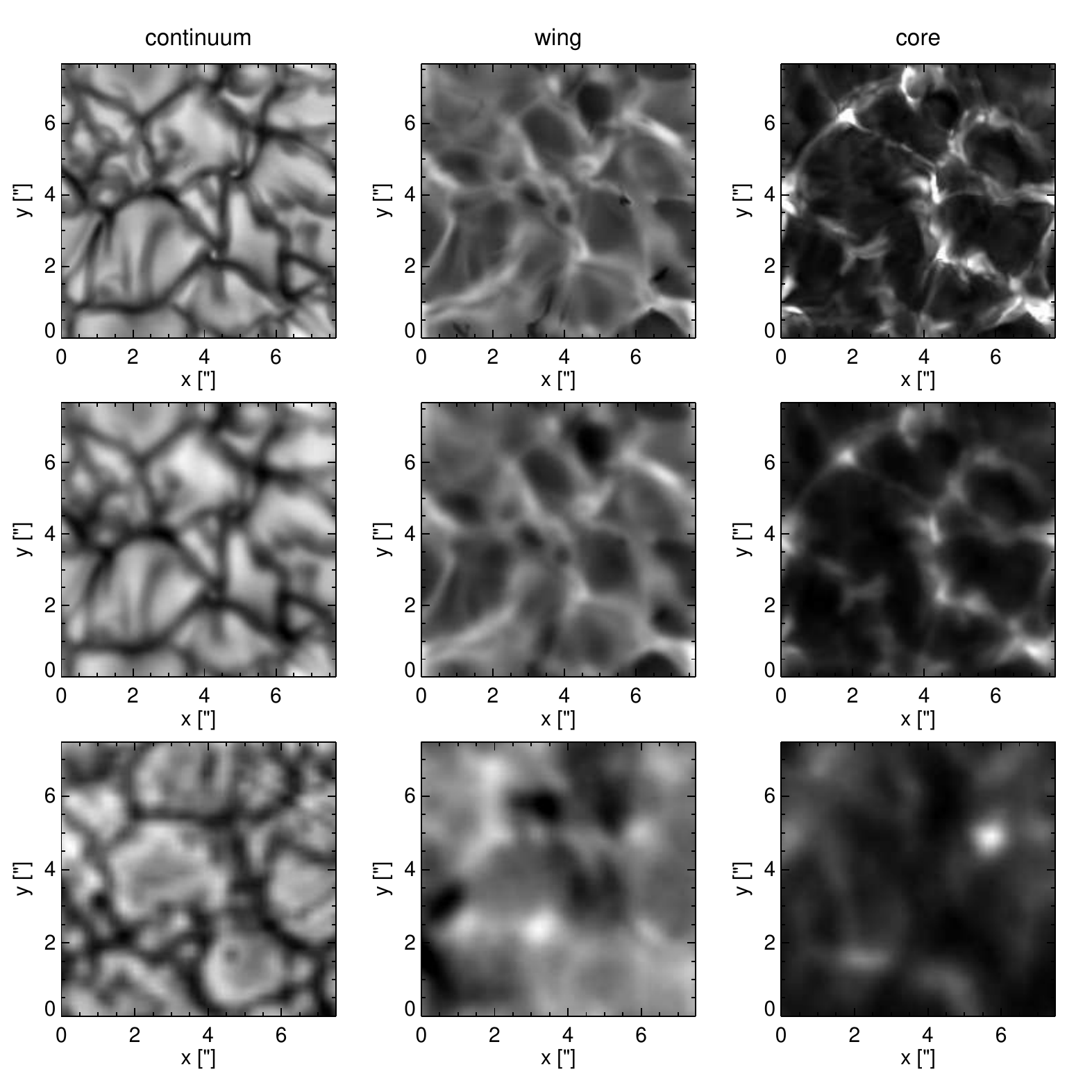}}
\caption{Small-scale structure of the solar atmosphere seen in the Ca~II
  infrared line at 854\,nm.  continuum (left column), line wing (middle), and
  line core (right column).  Top row: Synthetic maps based on a simulation
  with non-equilibrium hydrogen ionization; middle row: after application of a
  PSF and filter transmission; bottom: observations with IBIS at the DST
  (Courtesy of F.~W\"oger).  See text for details. }
\label{fig:swb_wca}
\end{figure}

Not only the modeling but also the observation of the shock-dominated layer
(hereafter referred to as ``fluctosphere'', see Sect.~\ref{sec:swb_atm}) is
non-trivial.  A clear detection in Ca~II~H, K or the infrared lines requires a
high spatial, temporal, and spectral resolution, all at the same time.  A too
broad filter wavelength range leads to significant contributions from the
photosphere below.  The fluctospheric pattern is then easily masked by a
reversed granulation signal.  The situation is complicated by the fact that
both patterns have very similar spatial scales, i.e.  roughly granulation
scales.  This is due to the fact that the generation of both patterns is due
to processes in the low photosphere.  This is illustrated in
Fig.~\ref{fig:swb_wca}, which shows preliminary synthetic intensity maps in
the Ca\,II infrared line at $\lambda = 854$\,nm.  The maps were calculated
with the non-LTE radiative transfer code MULTI \citep{Carlsson1986} column by
column from the model by \citet{2006A&A...460..301L}.  We use the
non-equilibrium electron densities, which are output from the time-dependent
simulation.  The top rightmost panel of Fig.~\ref{fig:swb_wca} shows the
mesh-like pattern in the line core, whereas the reversed granulation is
visible in the line wing (middle column).  Even further out in the wing, the
granulation pattern appears (left column).  The mesh-like fluctosphere pattern
can be seen Ca~H, K, and the IR triplet, too.

A comparison of the line core map with the temperature maps in
Fig.~\ref{fig:swb_xyslices} shows that primarily the hottest regions of the
pattern are seen in the Ca intensity, whereas a lot of atmospheric
fine-structure remains invisible.  The hot regions are caused by ``collision''
of neighboring shock fronts, ultimately compressing the gas in the region
in-between and rising its temperature.  This effect enhances in particular the
Ca brightness at the vertices of the mesh.  These small bright areas most
likely are observed as Ca~grains, while the emission along the mesh is so
faint that it is hard to detect.  The formation of Ca\,II~grains by
propagating shock waves was already explained by \citep{carlsson97a} over a
decade ago.  The fact that their detailed 1D simulations closely match
observations of grains, clearly shows that Ca\,II~grains are indeed a
phenomenon related to shock waves.  In 1D but also in 3D, the formation takes
place at heights of $\sim 1$\,Mm above optical depth unity.  In both cases,
the shocks propagate upwards into down-flowing material.  The difference,
however, is that in 1D shocks are plane-parallel so that
interaction between individual waves is essentially reduced to shock-merging
and shock-overtaking.  In 3D, shock wave interaction is more complex.  And
still, the compression zones between shocks -- the most likely candidate for
grain formation in 3D -- moves upwards with the waves and thus certainly show
very similar observational signatures.  While it seems to be well established
that Ca~grains are produced by shock waves, some details of the formation
process has to be revisited in a 3D context.

However, the grains might just be the ``tip of the iceberg''.  Progress in
observational techniques and instrumentation now finally allow us to detect
the dark details of the fluctosphere.  The middle row of
Fig.~\ref{fig:swb_wca} illustrates this observational effect.  A point spread
function (PSF) has been applied to the synthetic maps.  The PSF accounts for a
circular, unobstructed aperture of 70\,cm diameter and a non-ideal Voigt-like
contribution due to instrumental stray-light and atmospheric seeing.  Finally,
the degraded maps are integrated over wavelength with a synthetic transmission
filter with a FWHM of 5\,pm.  The assumptions are rather optimistic and
represent excellent observational conditions.  And yet the resulting image
degradation has a significant effect on the visible patterns.  Obviously, a
lower spatial or spectral resolution would further suppress the faint
mesh-like pattern in the line core.  Please note that the calculations are
still preliminary.  A full 3D treatment of the radiative transfer and the
included scattering, which will soon be possible, might increase the area of
enhanced brightness.  Also it is not clear yet how the possible interaction of
the shock waves with the overlying ``canopy'' field would alter the properties
of the pattern and its observational mesh/grain signature.  The resulting
pattern nevertheless in many aspects resembles the recent observations by
F.~W\"oger et al. with (i)~the InterferometricBIdimensional Spectrometer
\citep[][IBIS]{2007arXiv0709.2417C,2008A&A...480..515C} at the Dunn Solar
Telescope (DST) of the National Solar Observatory at Sacramento Peak
\citep{2008IAUS..247...66W} and (ii)~with the German Vacuum Tower Telescope
(VTT) at the Observatorio del Teide \citep{2006A&A...459L...9W}.  See the
lower row of Fig.~\ref{fig:swb_wca} for examples of IBIS data.

Based on the models by \citet{wedemeyer04a}, weak magnetic fields were
taken into account in the simulations by
\citet{2005ESASP.596E..65S,2006ASPC..354..345S} and
\citet{2008ApJ...680L..85S} (see Fig.~\ref{fig:swb_xyslices}).  Different
initial magnetic field configurations and strengths from $B_0 = 10$\,G to
20\,G were tried, all resembling quiet Sun internetwork conditions
\citep[see][ for an experiment with $B_0 = 100$\,G]{2005ESASP.596E..16W}.  The
computational domains again comprise several granules and extend into the
chromosphere, typically to heights of $\sim 1400$\,km.  The MHD models are
very similar to their hydrodynamic precursors with respect to structure and
dynamics.  The ubiquitous shock waves produce a very similar pattern in the
gas temperature but also shape the small-scale structure of the magnetic field
in the upper model atmosphere.  Consequently, the magnetic field in the
fluctosphere is highly dynamic and has a complex topology.  A look at
horizontal cross-sections at different heights in Fig.~\ref{fig:swb_xyslices}
implies that the field in the upper layers is much weaker ($|B| < 50$\,G) and
more homogenous than in the photosphere below .  On the other hand, the
fluctospheric field evolves much faster.  The horizontal field component
$B_\mathrm{hor}$ in the range 500\,km to 750\,km is (i)~stronger than the
vertical one, $B_\mathrm{z}$ and (ii)~has a rather large filling factor there.

\label{sec:plasmabeta}
In the small-scale internetwork simulations carried out with
\mbox{CO$^5$BOLD}, the strongly varying surface of plasma $\beta = 1$ is found
on average at heights of the order of $1000$\,km to $1400$\,km or even higher,
depending on model details.  Heights of the same order are also found by,
e.g., \citet{2007ASPC..369..193H}.  The exact location certainly depends on
the field strengths in the internetwork, which are still under debate.
Instead of plasma $\beta = 1$, one can also talk about an equivalent surface,
where sound speed and Alfv{\'e}n speed are equal.  It makes clear that these
regions are important for the propagation and eventual dissipation.
Simulations show that
this surface indeed separates two domains that differ in their dynamical
behavior: A slow evolving lower part and a highly dynamic upper part.  This
is certainly related to the finding that wave mode conversion and refraction
occurs under the condition of plasma $\beta \approx 1$
\citep{2002ApJ...564..508R,2003ApJ...599..626B,%
2007AN....328..286C,2007AN....328..323S}.  The current sheets, which are
present below and above the plasma $\beta = 1$ surface, differ in their
orientation.  While they are mostly stacked with horizontal orientation in the
lower part down to the top of the small-scale canopies at the boundary to the
photosphere, the thin current sheets above plasma $\beta = 1$ are formed along
shock fronts and can thus show oblique or even vertical orientation.

\subsection{Large-scale simulations}
\label{sec:swb_lss}

The models described in Sects.~\ref{sec:swb_inp} and \ref{sec:swb_inc} do not take 
into account the large-scale canopy fields but rather concentrate on the small
spatial scales of quiet Sun internetwork regions. 
In contrast, the simulations discussed in this section comprise larger
computational domains.  To make this possible, one usually has to make
compromises such as, e.g., reduce the spatial resolution or develop efficient
numerical methods.  \citet{2006ESASP.624E..79S} made impressive progress by
extending the computational box towards supergranulation scales.  Their models
do not include the upper atmosphere but extend deep into the convection zone.
\citet{2002ApJ...572L.113G,2005ApJ...618.1031G}, on the other hand, succeeded
in creating time-dependent numerical models, which extend from the photosphere
all the way into the corona.
An important aspect, which can be investigated with this kind of models, it the (magnetic) connection between the atmospheric layers all 
the way from the top of the convection to the corona. 
\citep[see also][]{2007ApJ...665.1469A}. 
Furthermore, extended simulations allow for investigating phenomena that are connected to spatial scales between granulation and supergranulation. 
For instance, the simulations by \citet{2005ESASP.592E..87H}
and \citet{2006ApJ...647L..73H} 
revealed the formation of dynamic chromospheric features similar to dynamic fibrils. 
Being driven 
by upward propagating waves in the chromosphere,  they are an example of the 
coupling between different atmospheric layers. 
Another type of coupling is provided in the form of horizontal magnetic flux structures with extensions of a few Mm, which emerge from the upper convection and rise upwards 
through the atmosphere. 
See \citet{2008ApJ...679..871M} and \citet{2007A&A...467..703C} for recent examples 
of flux emergence simulations. 
The simulations by \citet{2007A&A...473..625L} confirm once more \citep[cf.][]{carlsson02} that 
the ionization degree of hydrogen has to be treated in non-equilibrium in the upper atmosphere. 
Although the simulation is two-dimensional, it features weak-field sub-canopy domains with upward propagating shock waves and a magnetic-field dominated ``canopy'' domain above (see their Fig.~1). 
A strong coupling of the individual layers is very obvious.  

\begin{landscape}
\begin{figure}
\vspace*{-4mm}
\resizebox{\hsize}{!}{
\includegraphics*{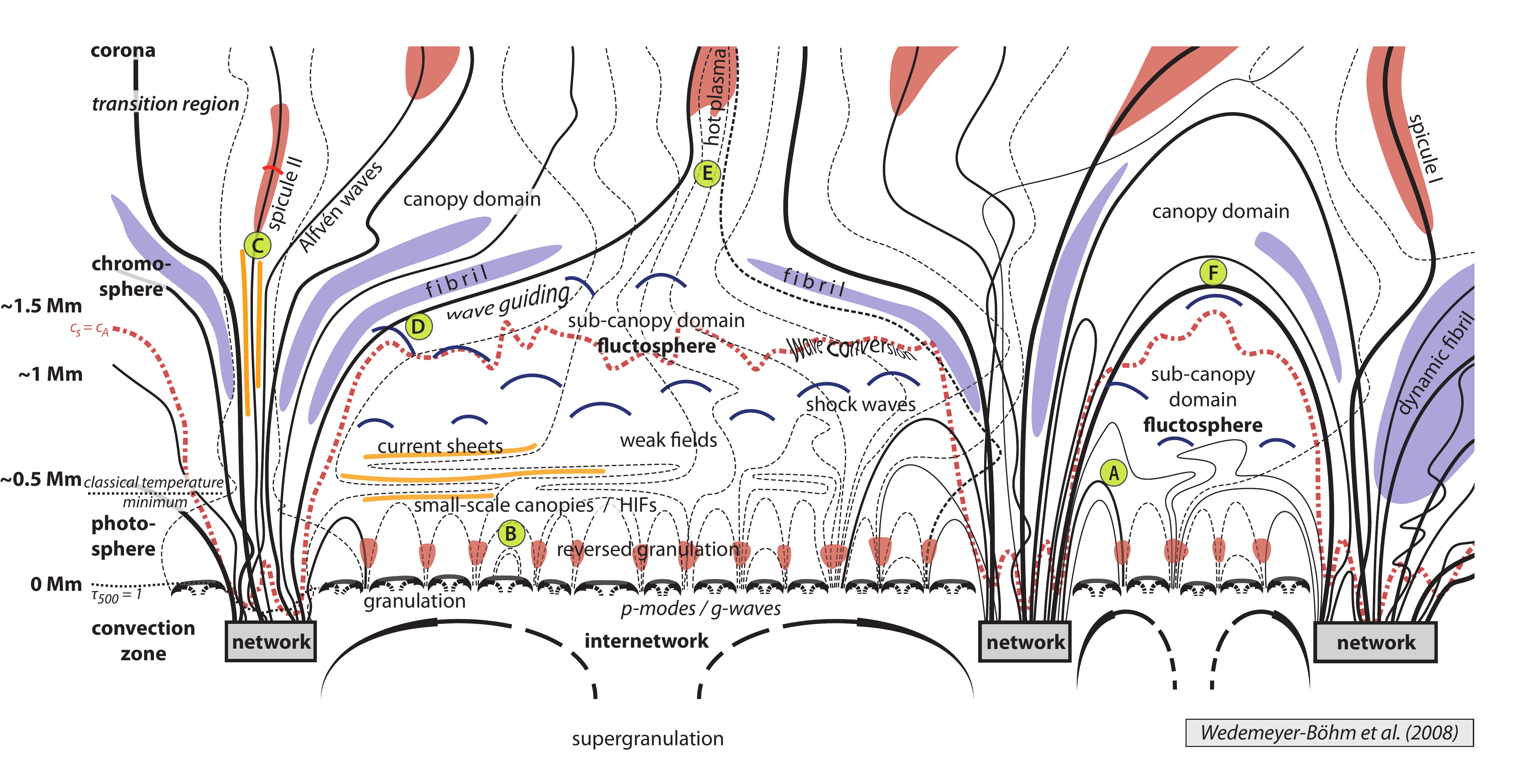}}
\vspace*{-7mm}
\caption{Schematic, simplified structure of the lower quiet Sun atmosphere
  (dimensions not to scale): The solid lines represent magnetic field lines
  that form the magnetic network in the lower layers and a large-scale
  (``canopy'') field above the internetwork regions, which ``separates'' the
  atmosphere in a canopy domain and a sub-canopy domain.  The network is found
  in the lanes of the supergranulation, which is due to large-scale convective
  flows (large arrows at the bottom).  Field lines with footpoints in the
  internetwork are plotted as thin dashed lines.  The flows on smaller spatial
  scales (small arrows) produce the granulation at the bottom of the
  photosphere ($z = 0$\,km) and, in connection with convective overshooting,
  the weak-field ``small-scale canopies''.  Another result is the formation of
  the reversed granulation pattern in the middle photosphere (red areas).  The
  mostly weak field in the internetwork can emerge as small magnetic loops,
  even within a granule~(point~B).  It furthermore partially connects to the
  magnetic field of the upper layers in a complex manner.  Upward propagating
  and interacting shock waves (arches), which are excited in the layers below
  the classical temperature minimum, build up the ``fluctosphere'' in the
  internetwork sub-canopy domain.  The red dot-dashed line marks a
  hypothetical surface, where sound and Alfv{\'e}n are equal.  The labels D-F
  indicate special situations of wave-canopy interaction, while location D is
  relevant for the generation of \mbox{type-II} spicules (see text for
  details).  Please note that, in reality, the 3D magnetic field structure in
  the canopy and also in the subcanopy is certainly more complex and entangled
  than shown in this schematic sketch.  }
\label{fig:swb_sketch}
\end{figure} 
\end{landscape}

\section{An updated picture of the quiet Sun atmosphere}
\label{sec:swb_atm}

The results of the previous sections are summarized in a schematic sketch
of the quiet Sun atmosphere (see Fig.~\ref{fig:swb_sketch}) with particular
emphasis on the low atmosphere in internetwork regions.  It is based on (and
should be interpreted in comparison with) recent sketches by, 
\citet{2006ASPC..354..259J}, and
\citet{2006ASPC..354..276R,2007ASPC..368...27R} but contains many
modifications to incorporate new results derived from observations and
numerical simulations.

\subsection{The large-scale magnetic field}

The large-scale building blocks of the quiet Sun atmosphere are the magnetic network
patches, which outline supergranulation cells.  The large-scale convective
flows (see long arrows) advect magnetic field to the lanes of the
supergranulation.  Consequently, the magnetic field is highly structured and
concentrated close to the ``surface'' ($\tau_{500} = 1$) with kG~field
strengths.  The visible result is the so-called magnetic network (see
Fig.~\ref{fig:wln_obs8}.  More recent observations with high spatial
resolution \citep[e.g.,][]{2007ApJ...670L..61O} reveal that the magnetic
network patches consist of a conglomerate of smaller magnetic elements or
``flux bundles'' of different field strength with a wealth of substructure.
This finding is incorporated in Fig.~\ref{fig:swb_sketch}, in contrast to
earlier sketches that feature the magnetic network as rather massive flux
tubes.  The heights where sound speed and Alfv{\'e}n are equal ($c_\mathrm{s}
= c_\mathrm{A}$ ), or equivalently where the plasma $\beta = 1$, will certainly show large
variations, depending on the (local) field strength.  It may even reach below
the surface of optical depth unity (at a reference wavelength of 500\,nm)
within strong field concentrations but may stay up at a few hundreds
kilometers in weaker network patches.  And still the variation in field
strength and topology, incl. the width of the network patches, is even larger
than can be presented in the simplified sketch here.

The magnetic field spreads out in the layers above the patches.  Depending on
the polarity of neighboring flux concentrations, they can form funnels or
connect via loops that span the internetwork regions in-between.  These two
cases are illustrated in Fig.~\ref{fig:swb_sketch} in a simplified way.  In
the classical picture, the large-scale field enclosing the weak-field
internetwork regions is referred to as ``magnetic canopy''.  The corresponding
flux funnels are often depicted with a wineglass-like geometry and have their
footpoints in the photospheric network only.  In reality, where the third
spatial dimension offers an important additional degree of freedom, the field
topology is more complex \citep[see,
  e.g.,][]{2006ASPC..354..331G,2006ApJ...638.1086P, 2006A&A...460..901J,
  2005ApJ...630..552S}.  \citet{2003ApJ...597L.165S} state that as much as
half of the field could actually be ``rooted'' in the internetwork regions.
From there, it can connect directly to the coronal field or via small loops to
the photospheric network.  The network patches could thus be surrounded by
``collars'' of loops with spatial scales comparable to one or a few granules.
Consequently, the concept of a regular canopy structure seems questionable.
Instead, the field topology should rather be understood as a set of individual
field lines.  Nevertheless, we stick here to the term ``canopy'' but use it in
a wider sense.  The height of the canopy and the field structure as a whole
varies significantly from region to region and with time.  The height
indicators to the left in Fig.~\ref{fig:swb_sketch} should therefore only be
used for rough orientation.  In principle, the lower boundary of the
``canopy'' field separates two distinct domains: a {\em canopy domain} and a
{\em subcanopy domain}.  In reality, however, the boundary is certainly less
strict than the sketch may imply.  Rather, the magnetic field of both domains
may be interconnected, e.g., by small loops, which extend on granular scales
(point~A).  This way, the dynamics of the internetwork photosphere could have
a direct influence on the properties of the upper layers, e.g., with respect to wave
propagation and heating.

\subsection{The canopy domain} 
\label{sec:canopydomain}

The canopy domain is dominated by (large-scale) magnetic fields.  It is this
layer, which, due to the emission in H$\alpha$, appears as a purple-red rim at
the beginning and end of a total solar eclipse.  Therefore, only the canopy
domain represents the chromosphere in a strict and original sense.  At a
closer look, a rich fibrilar structure can be seen in chromospheric H$\alpha$
observations.  They are found in rosette-like formations that funnel out from
the magnetic network below and in many cases connect to neighboring network
fields.  A few fibrils are shown in Fig.~\ref{fig:swb_sketch} in connection
with plasma that is trapped in the chromospheric field.  Such fibrils and also
the larger dynamic fibrils \citep[][shown at the right in the figure
  here]{2006ApJ...647L..73H,2008ApJ...673.1194L,2008ApJ...673.1201L} are an
integral part of the quiet Sun chromosphere and even more frequent than can be
shown in the 2D sketch here.
According to \citet{2007ASPC..368...65D}, fibrils could be the result of
chromospheric shock waves that occur when convective flows and global
oscillations leak into the chromosphere along the field lines of magnetic flux
concentrations.  In general, magnetohydrodynamic waves are an integral and
ubiquitous part of the canopy domain.  (Alfv{\'e}n waves are indicated in
Fig.~\ref{fig:swb_sketch} but represent just one of several possible wave modes).
Such perturbations can be excited by a number of processes, e.g., by the
shuffling and braiding of the magnetic footpoints in the photosphere by
convective flows.
As the large-scale magnetic field continues from the lower layers into the
{\em transition region} into the {\em corona} above, the whole canopy domain
is dynamically coupled.  Again, it must be emphasized that the field topology
is certainly more complex than can be expressed in the sketch here
\edt{(see, e.g., Fig.~\ref{fig:wln_obs8}).}
Indeed,
\citet{2005ApJ...630..552S} state that instead of the plasma-$\beta$ surface
being closely connected to the (classical) canopy, regions with low and high
$\beta$ can well be mixed up into the corona.

\edt{As already mentioned in Sect.~\ref{sec:observ}, a} 
most obvious constituent of the chromosphere, at least when observed at the
solar limb, are spicules \citep[see, e.g.,][]{2004Natur.430..536D}.  
\edt{Now, two types of spicules are  distinguished based on 
differences in their dynamic behavior \citep{2007PASJ...59S.655D}.  
}
Spicules
of type~I are the result of shock waves that are excited by disturbances in
the photosphere (e.g., in connection with p-modes) and propagate from there
along the magnetic field lines photosphere into the upper layers
\edt{\citep{2006ApJ...647L..73H,2007ApJ...660L.169R}.} 
Spicules of
type~II, on the other hand, are more dynamic but
thinner, exhibit higher velocities \edt{and} have shorter lifetimes
\edt{\citep[see, e.g.,][]{2008ApJ...679L.167L}. }
They are most likely generated by magnetic reconnection events. 
\edt{
Alfv{\'e}n waves, which by many are considered as an ubiquitous phenomenon in the upper atmosphere, 
can be detected in connection with spicules 
\citep{2007Sci...318.1574D}.}
An example is drawn in
the upper chromosphere above some vertically orientated current sheets
(point~C).

Another ingredient of the sketch are blobs of hot plasma in the corona,
although their exact position and shape needs further investigation.
\citet{2003ApJ...590..502D} showed that the emission is not correlated with
the centers of flux concentrations.  Rather, the emission seems to appear at
random locations.  
Although \citeauthor{2003ApJ...590..502D} refer to ``moss''
\citep{1999SoPh..190..409B,1999ApJ...520L.135F}, which is related to active
regions, there is no obvious reason why the situation should be different for the quiet Sun corona. 
Also, hot plasma
regions like the one marked with ``E'' in the sketch are certainly not
preferentially located directly above the middle of an internetwork region.
In reality, the entangled and skewed field topology will make such blobs -- if
existent in the way depicted here -- appear rather uncorrelated with the field
topology of the underlying magnetic network.

\subsection{The sub-canopy domain} 

The magnetic field in the sub-canopy domain is mostly weak \citep[see,
  e.g.,][]{2004Natur.430..326T,2007ApJ...670L..61O}, so that the plasma is larger
than one in the lower layers.  There, the field is essentially passively advected by
the hydrodynamic flow fields.  Convective motions and overshooting at the
``surface'' are the fundamental structuring agents, making the granulation the
dominant spatial scale.  Nevertheless, the weak fields in the subcanopy domain
most likely connect at least partially with the stronger canopy field.  This
feature is taken into account as integral part of the atmosphere sketch.
Unfortunately, the presentation remains rather speculative at this point as
many details of how and where the connections exactly take place are still
unknown.

Beside the magnetic field, the consequences of convective overshooting allow to 
divide the subcanopy domain into layers with distinct 
dynamics (from bottom to top): 
low photosphere, 
middle photosphere, 
high photosphere, 
fluctosphere. 

\paragraph{The lower and middle photosphere} 
exhibit the visible imprints of the solar surface convection.  The granulation
in the low photosphere is directly produced by small-scale convection cells
\citep[see, e.g.,][]{1990A&A...228..155N}, while the reversed granulation in
the middle photosphere is a second-order effect.  Gas is brought up by
convective overshooting in the granule interiors, adiabatically expanding and
cooling.  It streams down again in the intergranular lanes, where it is
compressed and heated.  In addition, p-modes, i.e. global oscillations, and
local acoustic events are important ingredients of the photospheric dynamics.
Recently, \citet{2008ApJ...681L.125S} presented new support for the idea that
gravity waves could play an important role, too.

The usually weak magnetic field is brought up from the convection zone below
and/or possibly locally generated by small-scale dynamo action close to the
surface.  In the photosphere, the weak field is more or less passively
advected towards the intergranular lanes but also towards the upper
photosphere.  The resulting field concentrations in the lanes become visible
as very small and confined structures, e.g., in G-band images \citep[see,
  e.g.,][for a recent example]{2008arXiv0806.0345D}.  In general, the
internetwork field in the photosphere exhibits significant inclination and
mixed polarity \citep[see, e.g.,][]{2008A&A...477..953M,2007ApJ...670L..61O}.
The granule interiors may become virtually field-free if there is no supply of
magnetic fields with the warm convective upflows.  Such voids are enclosed by
small-scale canopies.  Over most of the granulation, the horizontal field
component is stronger than the vertical.  This effect is observed as
``horizontal internetwork fields'' (HIFs).  Magnetic field can emerge also in
the form of small loops, which may have footpoints even within a granule (see
point~B in Fig.~\ref{fig:swb_sketch}).  This process, which was observed by
\citet{2007ApJ...666L.137C}, most likely adds to the accumulation of field
above granules.  In addition to emerging loops, \citet{2006ApJ...642.1246S}
report on flux that is submerging and thus disappears from the surface.

\paragraph{The upper photosphere} marks the boundary between the photosphere, which 
is controlled by the effects related to convective overshooting, and the
wave-dominated layer above.  This boundary can roughly be placed at the height
of the classical temperature minimum.  There, the temperature structure
appears smoothed out and less structured than above and below; it is here
that the average temperature amplitudes are smallest.  It is roughly the height
where the UV continuum at 160\,nm is formed (cf.  TRACE passbands).  The upper
photosphere is the layer, where the small-scale canopies have their top and
where stacked (horizontal) current sheet become most obvious.  This layer can
be seen as a kind of (dynamical) insulation between the internetwork
photosphere and fluctosphere.  This effect becomes obvious in simulations when
starting from an initial condition which feeds in field at the lower domain
boundary.  The photospheric field is built-up rather quickly but the field
above only after a time delay because it only slowly spreads into the strongly
subadiabatic stratification of the upper photosphere.

\paragraph{The fluctosphere:} 
\label{sec:fluctosphere}
The shock-dominated domain in subcanopy internetwork regions (see
Sect.~\ref{sec:swb_inc}), is referred to as ``fluctosphere'' by
\citet{2008IAUS..247...66W}, while \citet{2007ASPC..368...27R} uses the term
``clapotisphere''.  It is located between the photosphere and the part of the
chromosphere visible in H$\alpha$.  It is composed of propagating and
interacting shock waves (with weak field only) and intermediate cool
post-shock regions.  Ideally, the wave fronts would expand spherically, while
moving in vertical direction.  In reality, they are deformed by running into
an inhomogeneous medium of downflowing gas, which was shaped by precursory wave
trains.  The horizontal expansion of the fronts inevitably causes interaction
between them.  A visible result is the formation of Ca grains at heights,
which traditionally would be assigned to the low chromosphere.  The waves are
excited in the photosphere below via different processes, which are related to
convection (e.g., exploding granules), overshooting, and p-modes.  The
magnetic field in the fluctosphere is rather weak and is therefore mostly
passively shuffled around by the shock waves.  The result is a very dynamic
and entangled field.  The strongly varying surface of plasma $\beta = 1$ or in
this context better $c_\mathrm{s} = c_\mathrm{A}$ is most likely located at
heights of the order 1000\,km to 1500\,km or even higher (see
Sect.~\ref{sec:plasmabeta}).  There, the conditions allow wave mode
conversion, so the parts of the fluctosphere below and above can show a
somewhat different dynamical behavior.  In the upper part, the weak fields
become more important and rapidly moving filaments of enhanced field strength
are generated.  The propagating shock waves nevertheless remain the dominating
structuring agent.  A consequence, however, is that the current sheets are
only stacked at plasma $\beta > 1$.  Above, they are less regular as they are
formed in the narrow collision zones of shocks, where the magnetic field is
occasionally compressed.  This shock-induced magnetic field compression might
qualify as a (minor) heating process with potential consequences for the
chromospheric energy balance.

The fluctosphere is not directly visible in H$\alpha$ (in the line core at
least) and is thus not a part of the chromosphere in a strict sense.  It seems
advisable to reserve the term chromosphere for the fibrilar canopy domain as
visible in H$\alpha$ (or in the very line cores of the Ca\,II lines).
However, the fluctospheric shock waves could still leave an imprint in
chromospheric diagnostics by interacting and penetrating the canopy field.
On the other hand, the fluctosphere is also no part of the photosphere,
although causally connected via the shock waves that propagate upwards from
the low photosphere.  The fluctosphere could be regarded as a second-order
effect only, in contrast to the granulation and reversed granulation, which
are direct consequences of the solar surface convection.

\subsection{Shock waves meet the ``canopy''} 

Some details of Fig.~\ref{fig:swb_sketch} concern the interplay of propagating
waves and the magnetic canopy.  There are certain zones in these magnetic
structures that act as mode conversion zone
\citep{2003ApJ...599..626B,2007AN....328..286C}, e.g., converting incoming
acoustic waves into other modes, such as fast and slow magnetoacoustic
waves.  It is thus possible that such converted waves continue to propagate
along the canopy field lines as some kind of ``canopy waves''.
For simplicity, such a zone is marked by ``wave conversion'' at the ``outer''
boundary between canopy and subcanopy domain in the figure.  Generally, such
zones can be located everywhere in the structure where sound speed and
Alfv{\'e}n speed are of equal magnitude.

Furthermore, refraction and even reflection of waves can occur in such zones.
As for the mode conversion, details depend on the relative orientation of the
incoming wave and the magnetic field \citep{2008IAUS..247...78H}.  A wave can
remain barely affected by the field when traveling perpendicular to the field
lines, e.g., upwards in a vertical flux concentration.  On the other hand,
significant (relative) inclination can even result in total internal
reflection for some wave modes \citep{2002ApJ...564..508R}.
In general, it can be assumed that the (acoustic) shock waves coming from the
fluctosphere are guided by the magnetic canopy (e.g., point~D in
Fig.~\ref{fig:swb_sketch}).  Consequently, waves might follow the canopy field
upwards and compress and heat the gas trapped between chromospheric
``funnels'' (point~E).  In closed loop regions, strong waves could push into
the canopy from below and compress the magnetic field (location F).  Depending
on the local field configuration and the properties of the incoming wave, such
an event could eventually trigger reconnection events.  It could contribute to
chromospheric heating.  It certainly would not be only limited to the
locations indicated in the sketch but occur more often in complex 3D field
configurations.  On the other hand, a regular closed structure as in the
figure could possibly refract the waves from below such that they are
``focussed'' in the top of the subcanopy domain, amplifying their effect on
the canopy field.  A possible -- although speculative -- result could be the
triggering of ``nanoflares'', although they are initiated by other mechanisms
at other locations, too.
That the upwards propagating waves interact with the canopy field is implied
by observations in H$\alpha$.  The dynamic behavior of the chromospheric
fibrils is reminiscent of strings that sway back and forth in reaction to the 
quasi-continuous impact of waves from below (point~F).  Under certain conditions, the
shock waves might actively deform the field configuration of the magnetic
canopy.  Magnetoacoustic waves can already enter network flux concentrations
in the photosphere, where the inclined magnetic field lines act as
``magnetoacoustic portals'' \citep{2006ApJ...648L.151J}.  The observations of
so-called ``acoustic shadows'' provide observational evidence for the
interaction of acoustic waves with the field around network footpoints
\citep{2001A&A...379.1052K,2001ApJ...548L.237M,2001ApJ...561..420M}.

\subsection{Probing the upper atmosphere}

With the currently available diagnostics for the chromosphere, observations of
the subcanopy domains are problematic.  
The H$\alpha$ line core samples only the ``canopy domain'', whereas observations 
in the line wing reveal a background that most likely is dominated by the 
reversed granulation at much lower heights.  
It seems questionable if the layer in-between -- the fluctosphere -- can be 
observed in the H$\alpha$ line wing at all in internetwork regions.  
Polarimetric measurements in the He\,10830\,\AA{} line (see Sect.~\ref{sec:observ}) 
principally allow for the determination of the magnetic structure in a slab located between 
1000\,km and 2000\,km. 
The formation of this line requires coronal illumination in the UV, resulting in 
complete absence of any photospheric contamination. 
However, the main contribution in the He\,10830\,\AA{} line comes from layers slightly 
above the fluctosphere. 
The Ca H \& K and IR lines in principle would allow observations of the fluctosphere 
if very narrow filters are used.  
Otherwise, the detected intensity is ``contaminated'' with radiation from layers below.  
Very often, Ca~observations with too broad filter prominently show the reversed
granulation (see Figs.~\ref{fig:wln_obs8}d and \ref{cak_fibrils_label.pdf}), 
which is easily mistaken as chromospheric signal.  
Very narrow filters, on the other hand, make it necessary to properly correct 
for Doppler shifts.  
A solution is fast scans through the Ca~II IR lines with new imaging 
polarimeters such as IBIS
\citep{2007arXiv0709.2417C,2008A&A...480..515C,kleint:08a}
or CRISP
\citep{2008arXiv0806.1638S}, 
or spectro-polarimeters like SPINOR
\citep[]{socasnavarro:06a}.  
The extended formation height ranges and the non-equilibrium conditions, under 
which the inner parts of these lines are formed, complicate the interpretation 
and the derivation of the atmospheric structure.  
A promising alternative are the (sub-)millimeter continua, which will become  
accessible with the Atacama Large Millimeter Array (ALMA) a few years
from now.  
Although technical details of this new type of observation render the 
construction of brightness temperature maps a certainly very complicated
task, the scientific results could significantly contribute to our
understanding of the solar atmosphere at chromospheric heights
\citep{2008Ap&SS.313..197L,2007A&A...471..977W}. 

\section{Conclusions}

The solar atmosphere is a very dynamic and inhomogeneous multi-scale system. 
Its individual components are coupled; some of them even show 
a kind of hierarchical self-similarity. 
Examples are the observational imprints of sub-surface convection, with
a continuous spectrum of scales from below granulation scales to above 
supergranulation scales, and magnetic fields, which
also exhibit similar features over a large range of spatial scales. 
 
Despite great progress on the theoretical and observational sides, 
which go hand in hand, we are still missing an ultimate, comprehensive 
picture of the quiet Sun atmosphere. 
But at least we can now see what is needed for a corresponding numerical 
simulation.  
First, the computational domain should be large enough to encompass a few 
supergranulation cells while the spatial resolution must still be high enough 
to capture important processes that occur on scales smaller than granulation. 
The vertical couplings make it necessary to consider an extensive height range.
The corona and chromosphere can only be treated realistically when including 
the important driving motions in the layers below, i.e. in the photosphere 
and (at least) the upper part of the convection zone. 
While many simplifying assumptions can be made for the lower parts of such a 
model, the layers above the (middle) photosphere require a numerically 
complicated and thus computationally expensive non-equilibrium modeling approach, 
e.g., a realistic treatment of hydrogen ionization etc..
The production of such a comprehensive model -- and analogous models for, e.g., 
active regions -- is thus very involved and can 
be regarded as one of the current challenges in (computational) solar physics. 

On the observational side, we must continue to push forward the 
instrumental possibilities towards higher resolution in the spatial, temporal, and spectral
domains -- all at the same time. 
In addition, the (further) development and exploitation of advanced diagnostics
is needed to derive a seamless tomography of the atmosphere as an integral phenomenon. 

Until we succeed to reach these ambitious goals, we are left with a number of 
open questions.  
Of particular interest for the quiet Sun are, amongst others: 

\begin{itemize}
\item How does the weak internetwork field connect with the stronger network field? 
 What does the magnetic field look like just below the ``canopy''? 
 And can we talk about a ``canopy'' even in a wider sense after all? 
\item How do the propagating fluctospheric shock waves interact with the stronger 
(``canopy'') field?  
Is it mostly a ``passive'' refraction / reflection at the ``boundaries'' of 
flux concentrations
\citep[e.g.][]{2002ApJ...564..508R, 2007AN....328..323S} 
or also ``active'' distortion/displacement/compression of magnetic field? 
How and where does mode conversion take exactly place under realistic conditions? 
\item The question of the coupling between the atmospheric layers is closely connected
  to the heating mechanism question or, better said, to the question of the 
  atmospheric energy balance, not only for the Sun but also for other stars. 
  Amongst other things, the ongoing controversy concerning the heating mechanism 
  of the quiet Sun chromosphere 
  \citep[e.g.,][]{2005Natur.435..919F} has 
  important implications for stellar activity in general. 
\end{itemize}

\textbf{\ \\ \noindent{}Acknowledgments:} 
We would like to thank O.~Steiner, K.~Schrijver, B.~de~Pontieu, M.~Carlsson, 
V.~Hansteen, R.~Rutten, and H.~Peter for helpful discussions. 
SWB was supported with a 
Marie Curie Intra-European Fellowship of the European Commission 
 (6th Framework Programme, FP6-2005-Mobility-5, Proposal No. 042049).   
{\AA}N acknowledges support from the Danish Natural Science Research Council 
and from the Danish Center for Scientific Computing.

\bibliographystyle{aa} 
\bibliography{p08issi_sect8_v_ref1}

\end{document}